\documentclass[letterpaper,american,reprint, aps, prl, superscriptaddress]{revtex4-1}
\usepackage[T2A,LGR,T1]{fontenc}
\usepackage[latin9]{inputenc}
\usepackage{array}
\usepackage{float}
\usepackage{booktabs}
\usepackage{textcomp}
\usepackage{amsmath}
\usepackage{amssymb}
\usepackage{graphicx}

\makeatletter

\pdfpageheight\paperheight
\pdfpagewidth\paperwidth

\DeclareRobustCommand{\greektext}{%
  \fontencoding{LGR}\selectfont\def\encodingdefault{LGR}}
\DeclareRobustCommand{\textgreek}[1]{\leavevmode{\greektext #1}}
\ProvideTextCommand{\~}{LGR}[1]{\char126#1}

\DeclareRobustCommand{\cyrtext}{%
  \fontencoding{T2A}\selectfont\def\encodingdefault{T2A}}
\DeclareRobustCommand{\textcyr}[1]{\leavevmode{\cyrtext #1}}

\providecommand{\tabularnewline}{\\}

\def \mpx {MPX$_3$}
\def \mps {MPS$_3$}
\def \feps {FePS$_3$}

\def \mnps {MnPS$_3$}
\def \vps {VPS$_3$}
\def \vpps {V$_{0.9}$PS$_3$}
\def \vpsevenps {V$_{0.78}$PS$_3$}

\makeatother

\usepackage{babel}
\begin{document}
\title{Isostructural Mott Transition in 2D honeycomb antiferromagnet \vpps{}}
\author{M.J. Coak}
\affiliation{Center for Correlated Electron Systems, Institute for Basic Science,
Seoul 08826, Republic of Korea}
\affiliation{Department of Physics and Astronomy, Seoul National University, Seoul
08826, Republic of Korea}
\affiliation{Department of Physics, University of Warwick, Gibbet Hill Road, Coventry
CV4 7AL, UK}
\affiliation{Cavendish Laboratory, University of Cambridge, J.J. Thomson Ave, Cambridge
CB3 0HE, UK}
\author{S. Son}
\affiliation{Center for Correlated Electron Systems, Institute for Basic Science,
Seoul 08826, Republic of Korea}
\affiliation{Department of Physics and Astronomy, Seoul National University, Seoul
08826, Republic of Korea}
\author{D. Daisenberger}
\address{Diamond Light Source, Chilton, Didcot OX11 0DE, UK}
\author{H. Hamidov}
\address{Cavendish Laboratory, University of Cambridge, J.J. Thomson Ave, Cambridge
CB3 0HE, UK}
\address{Navoiy Branch of the Academy of Sciences of Uzbekistan, Galaba Avenue,
Navoiy, Uzbekistan}
\affiliation{National University of Science and Technology \textquotedblleft MISiS\textquotedblright ,
Leninsky Prospekt 4, Moscow 119049, Russia}
\author{C.R.S Haines}
\address{Department of Earth Sciences, University of Cambridge, Downing Street,
Cambridge CB2 3EQ, UK}
\address{Cavendish Laboratory, University of Cambridge, J.J. Thomson Ave, Cambridge
CB3 0HE, UK}
\author{P.L. Alireza}
\address{Cavendish Laboratory, University of Cambridge, J.J. Thomson Ave, Cambridge
CB3 0HE, UK}
\author{A.R. Wildes}
\address{Institut Laue-Langevin, CS 20156, 38042 Grenoble C{\'e}dex 9, France}
\author{C. Liu}
\affiliation{Cavendish Laboratory, University of Cambridge, J.J. Thomson Ave, Cambridge
CB3 0HE, UK}
\author{S.S. Saxena}
\affiliation{Cavendish Laboratory, University of Cambridge, J.J. Thomson Ave, Cambridge
CB3 0HE, UK}
\affiliation{National University of Science and Technology \textquotedblleft MISiS\textquotedblright ,
Leninsky Prospekt 4, Moscow 119049, Russia}
\author{Je-Geun Park}
\affiliation{Center for Correlated Electron Systems, Institute for Basic Science,
Seoul 08826, Republic of Korea}
\affiliation{Department of Physics and Astronomy, Seoul National University, Seoul
08826, Republic of Korea}
\date{\today}
\begin{abstract}
We present the observation of an isostructural Mott insulator-metal
transition in van-der-Waals honeycomb antiferromagnet \vpps{} through
high-pressure x-ray diffraction and transport measurements. The \mpx{}
family of magnetic van-der-Waals materials (M denotes a first row
transition metal and X either S or Se) are currently the subject of
broad and intense attention, but the vanadium compounds have until
this point not been studied beyond their basic properties. We observe
insulating variable-range-hopping type resistivity in \vpps{}, with
a gradual increase in effective dimensionality with increasing pressure,
followed by a transition to a metallic resistivity temperature dependence
between 112 and 124~kbar. The metallic state additionally shows a
low-temperature upturn we tentatively attribute to the Kondo Effect.
A gradual structural distortion is seen between 26-80~kbar, but no
structural change at higher pressures corresponding to the insulator-metal
transition. We conclude that the insulator-metal transition occurs
in the absence of any distortions to the lattice - an isostructural
Mott transition in a new class of two-dimensional material, and in
strong contrast to the behavior of the other \mpx{} compounds.
\end{abstract}
\maketitle
Layered two-dimensional van-der-Waals materials are currently the
subject of broad and detailed research \citep{Ajayan2016}. In particular,
the addition of magnetism into such systems leads to many interesting
fundamental questions and opportunities for device applications \citep{Park2016,Samarth2017,Zhou2016,Burch2018},
and the ability to select or tune electronic and transport properties
in these materials would be a powerful tool indeed for the fabricators
of a new generation of nanoscale devices. One particular family of
materials enjoying a sudden surge of interest is that of \mpx{},
where M denotes a first row transition metal and X either S or Se.
First synthesized by Klingen in 1969 \citep{Klingen1968,Klingen1970,Klingen1973},
initial interest in these materials beyond their basic characterization
was for application as battery materials - see Grasso and Silipigni
\citep{Grasso2002} for a review. In more recent years they have been
studied in detail as excellent examples of two-dimensional magnetic
systems - these materials all share very similar structures, but spin
states, magnetic ordering, magnetic anisotropy and critical behavior
change with the transition metal \citep{Kurosawa1983,Okuda1986,Wildes1994,Wildes1998a,Rule2007,Wildes2007,Wildes2012,Wildes2015,Wildes2017,Lancon_2016}.
\mpx{} form a layered honeycomb lattice of the metal ions \citep{Brec1979,Ouvrard1985,Ouvrard1985a,Brec1986}
with monoclinic space group $C2/m$ and interplanar forces solely
through a van-der-Waals interaction between the surrounding P$_2$S$_6$
clusters. They can be easily mechanically exfoliated as with graphene
and have been shown to maintain their magnetic ordering down to monolayer
thickness \citep{Lee_2016,Kuo2016}. These materials are all insulating
- they exhibit an exponentially increasing resistivity with decreasing
temperature - and can be understood as p-type semiconductors \citep{Grasso2002}
and as Mott insulators \citep{Haines2018b}. Recent works have demonstrated
Mott insulator-metal transitions in \mnps{} and \feps{} \citep{Wang2016a,Tsurubayashi2018,Haines2018b}
and additionally superconductivity in FePSe$_3$ \citep{Wang2018}.
The tuning of clean and controllable materials like these from an
antiferromagnetic Mott insulating state into a metallic, or indeed
superconducting, state is of great interest for fundamental magnetism
and Mott physics. Moreover, this same physics forms the foundation
for our understanding of the underlying phase diagram and mechanisms
for systems like the cuprate superconductors.

\vps{}, or more generally V$_{1-x}$PS$_3$ with $x$ the level of
vanadium deficiency, is a member of the family that has received very
little attention, despite hosting great potential for interesting
study. It has the smallest band gap of these insulating materials
at around 0.25~eV \citep{Brec1980,Ouvrard1985a,Ichimura1991}, and
by far the lowest resistivity (on the order of $\Omega$cm) at room
temperature. This can additionally be tuned over an order of magnitude
by altering the level of vanadium deficiency \citep{Ichimura1991}.
Theoretical band structure calculations for the whole material family,
including \vps{} are given in Chittari et.al. \citep{Chittari2016},
showing a band gap and insulating/semiconducting behavior, but such
calculations on these materials are challenging and historically often
contradict with experiment. V$_{1-x}$PS$_3$ is antiferromagnetic
\citep{Ouvrard1985a}, with a N{\'e}el temperature of around 62~K but
little is known about its magnetic structure and behavior. As the
metal ion in \mpx{} must take the charge M$^{2+}$, the vanadium
deficiency in V$_{1-x}$PS$_3$ can be explained as due to valence
mixing on the vanadium site between V$^{2+}$ and V$^{3+}$ states.
It is this valence mixing that Ichimura and Sano \citep{Ichimura1991}
argue to be responsible for the comparatively high conductivity in
this material - but the resulting high degrees of local electronic
disorder and vacancies can be expected to have a large effect on the
transport and scattering properties.

\section*{Results}

\subsection*{Crystal structure}

\begin{figure}
\begin{raggedright}
a)\hspace{0.5\columnwidth}b)
\par\end{raggedright}
\begin{centering}
\includegraphics[width=0.5\columnwidth]{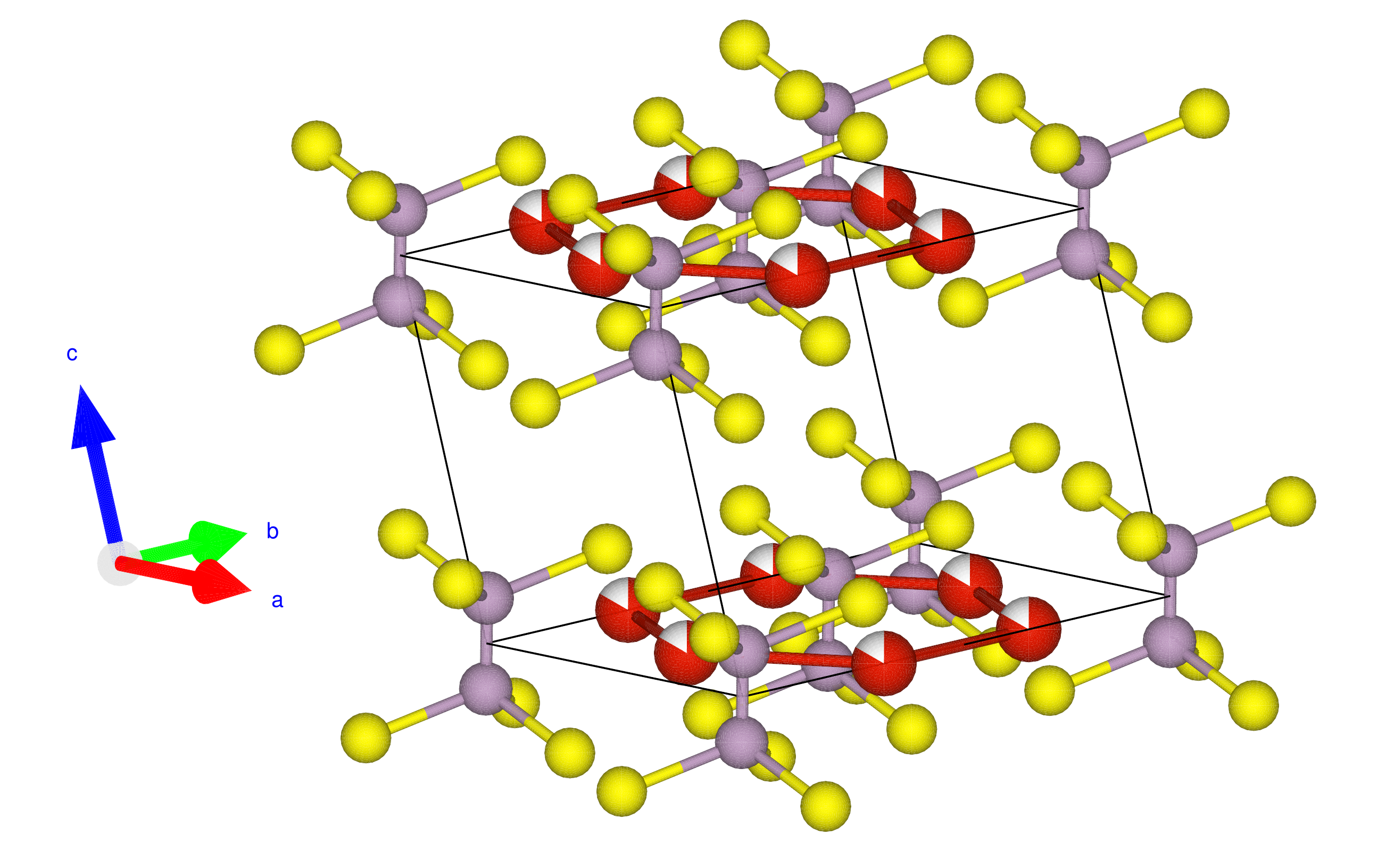}\includegraphics[width=0.5\columnwidth]{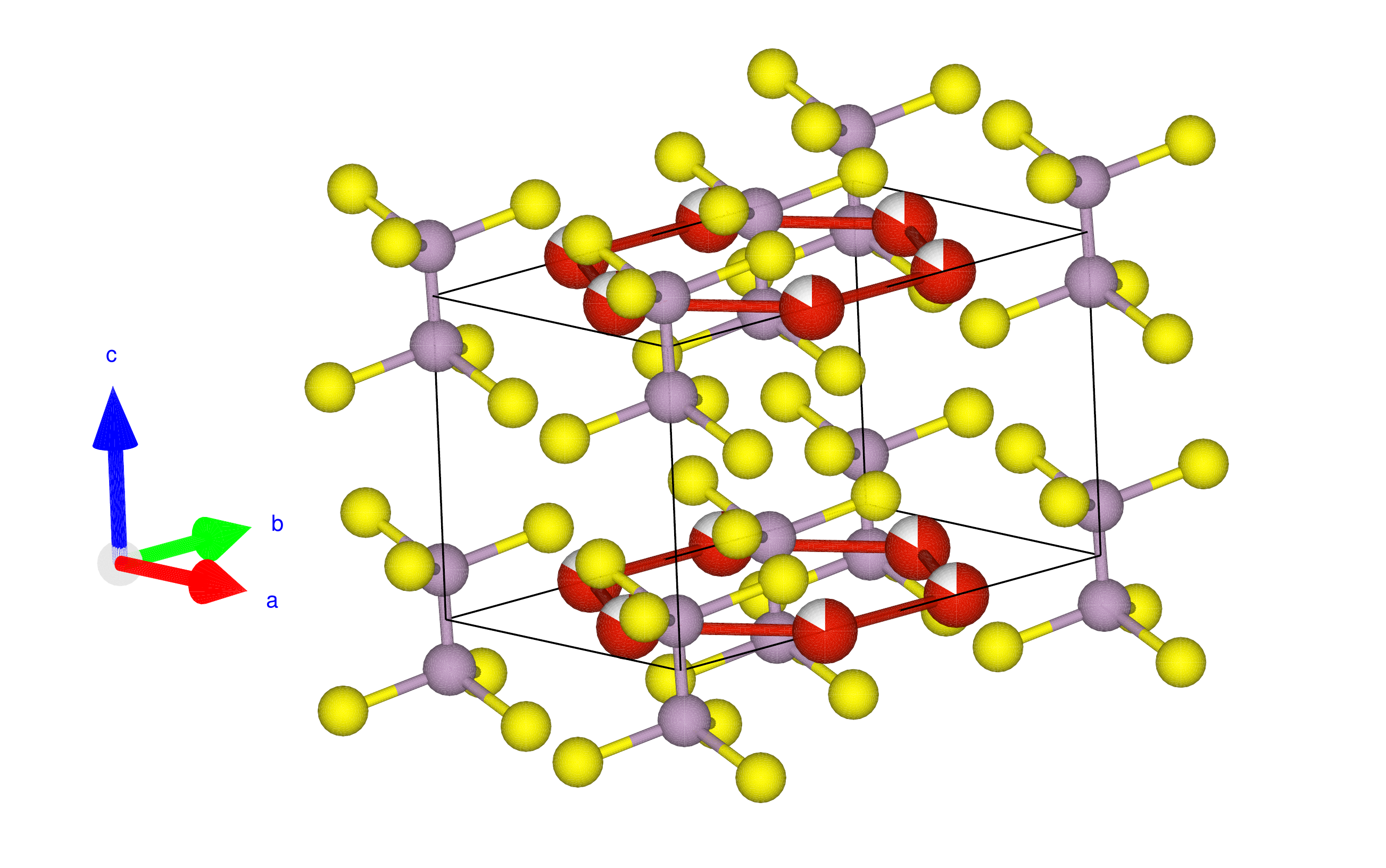}
\par\end{centering}
\begin{raggedright}
c)\hspace{0.5\columnwidth}d)
\par\end{raggedright}
\begin{centering}
\includegraphics[width=0.45\columnwidth]{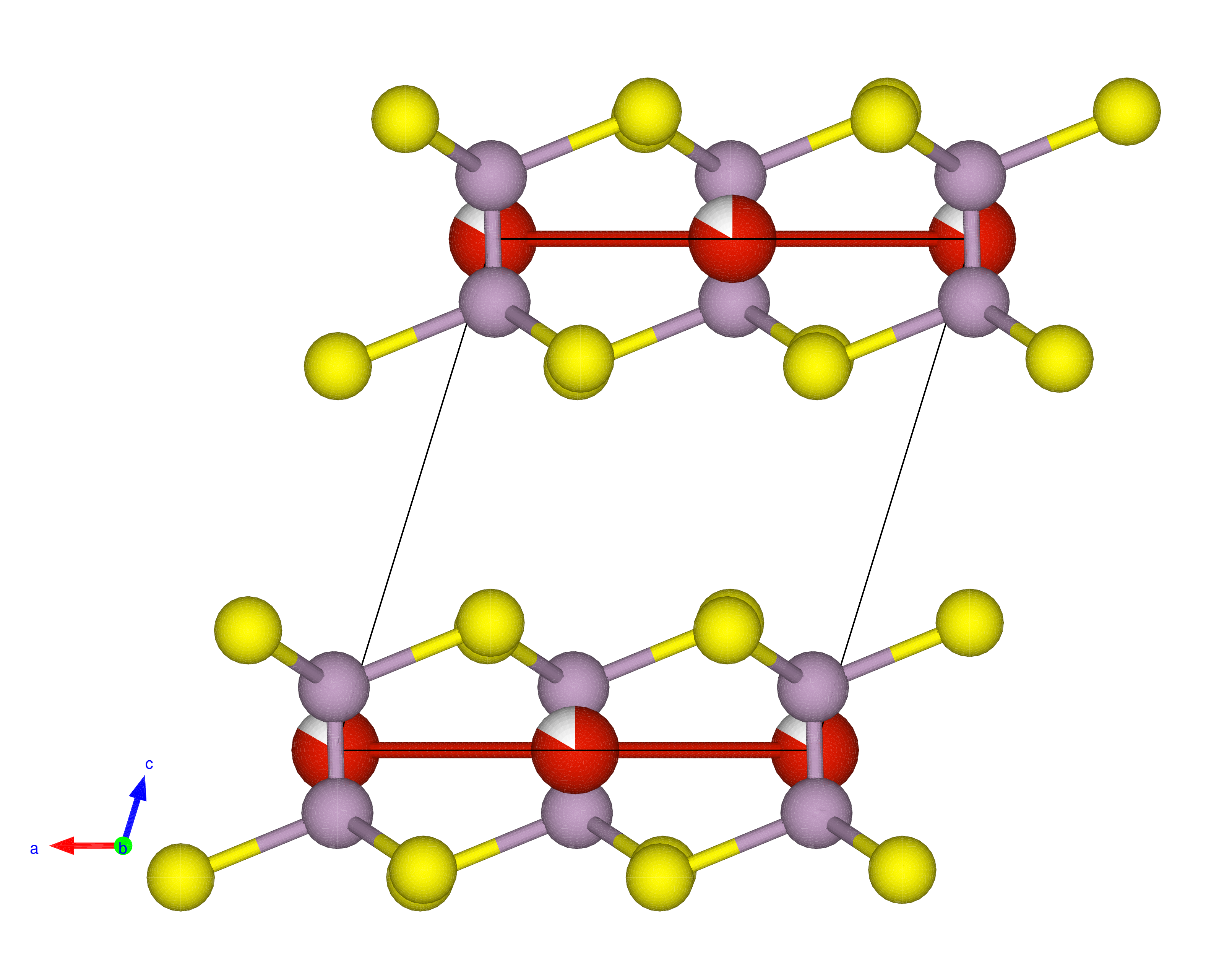}\includegraphics[width=0.45\columnwidth]{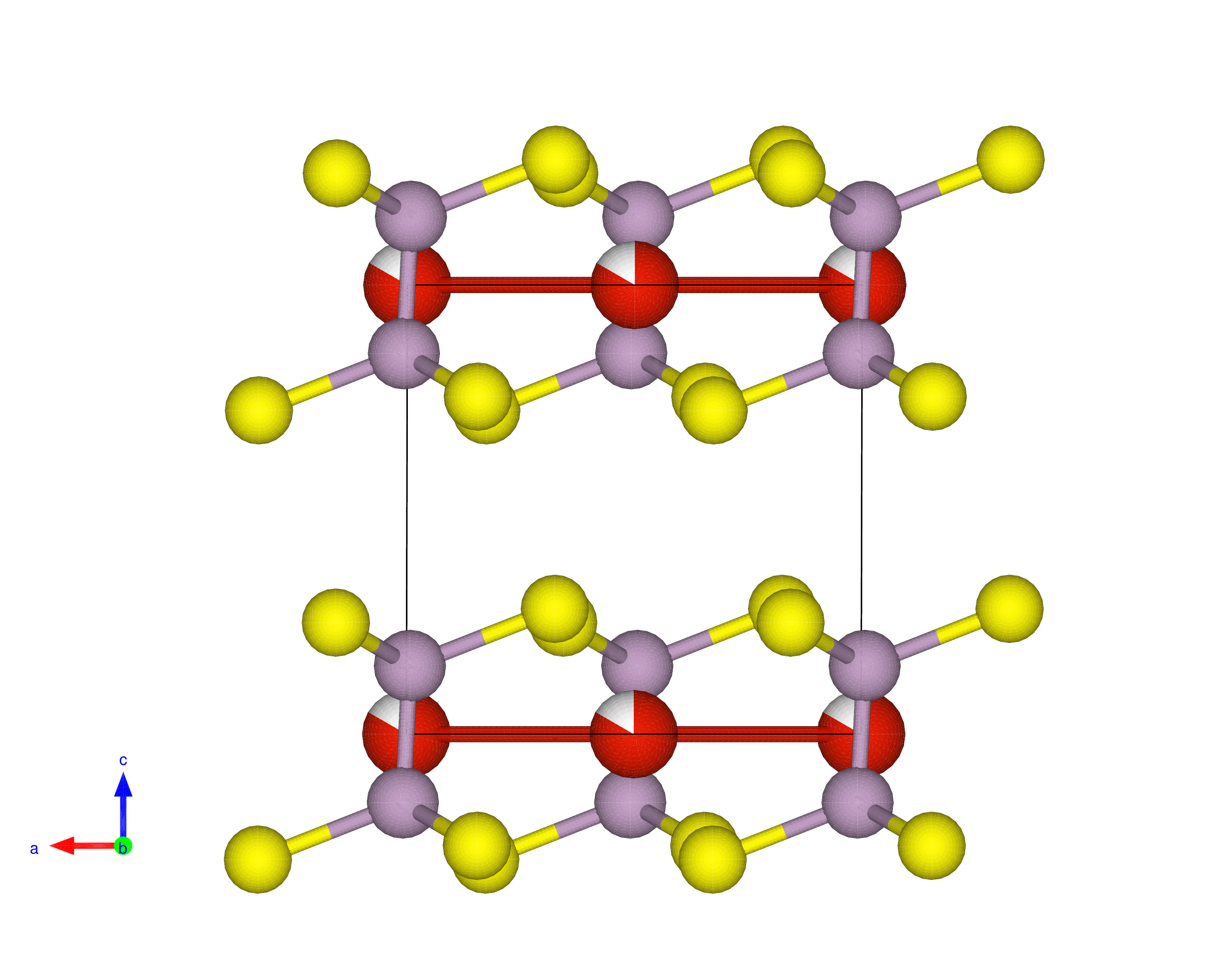}
\par\end{centering}
\begin{raggedright}
e)\hspace{0.5\columnwidth}f)
\par\end{raggedright}
\centering{}\includegraphics[width=0.5\columnwidth]{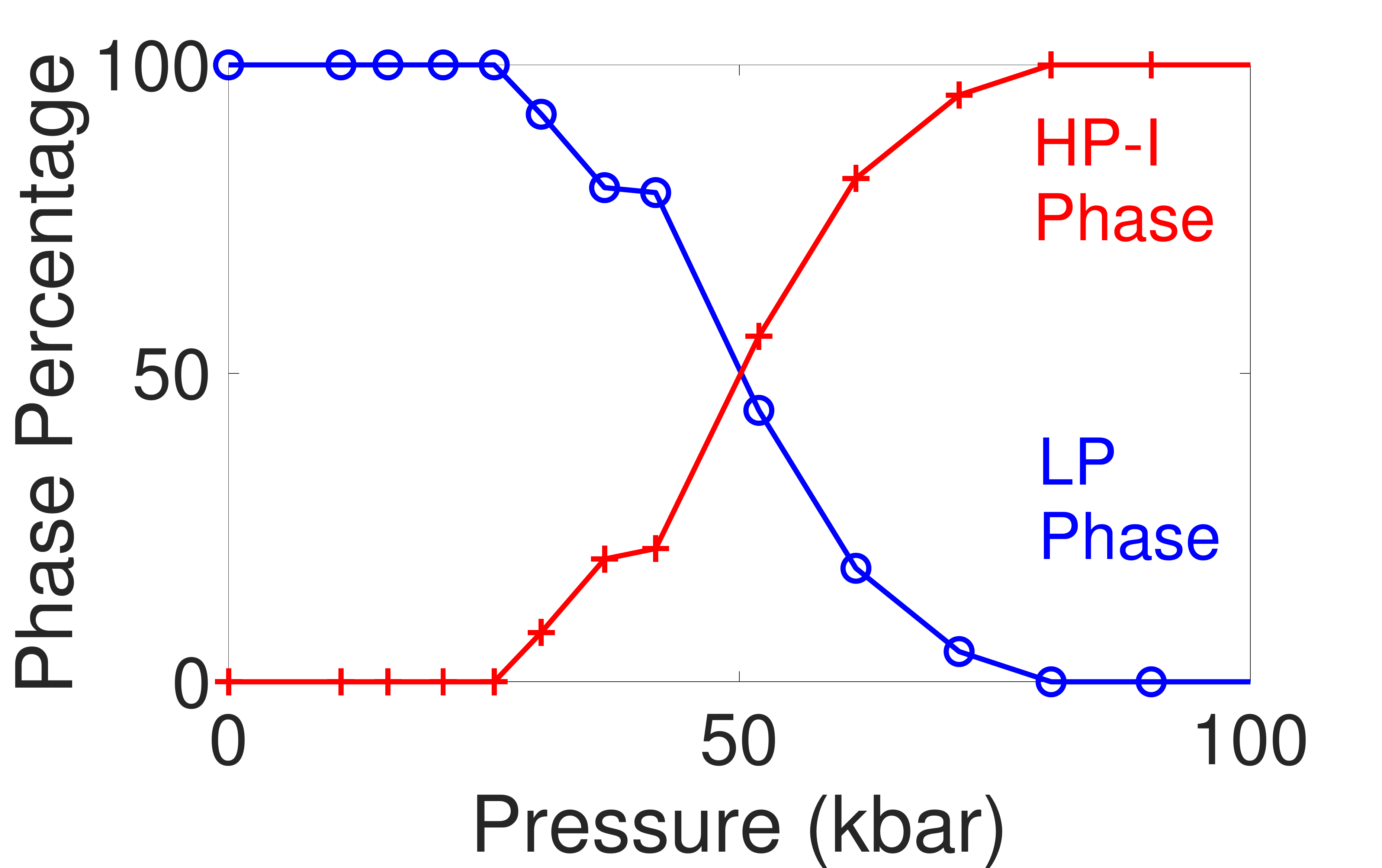}\includegraphics[width=0.5\columnwidth]{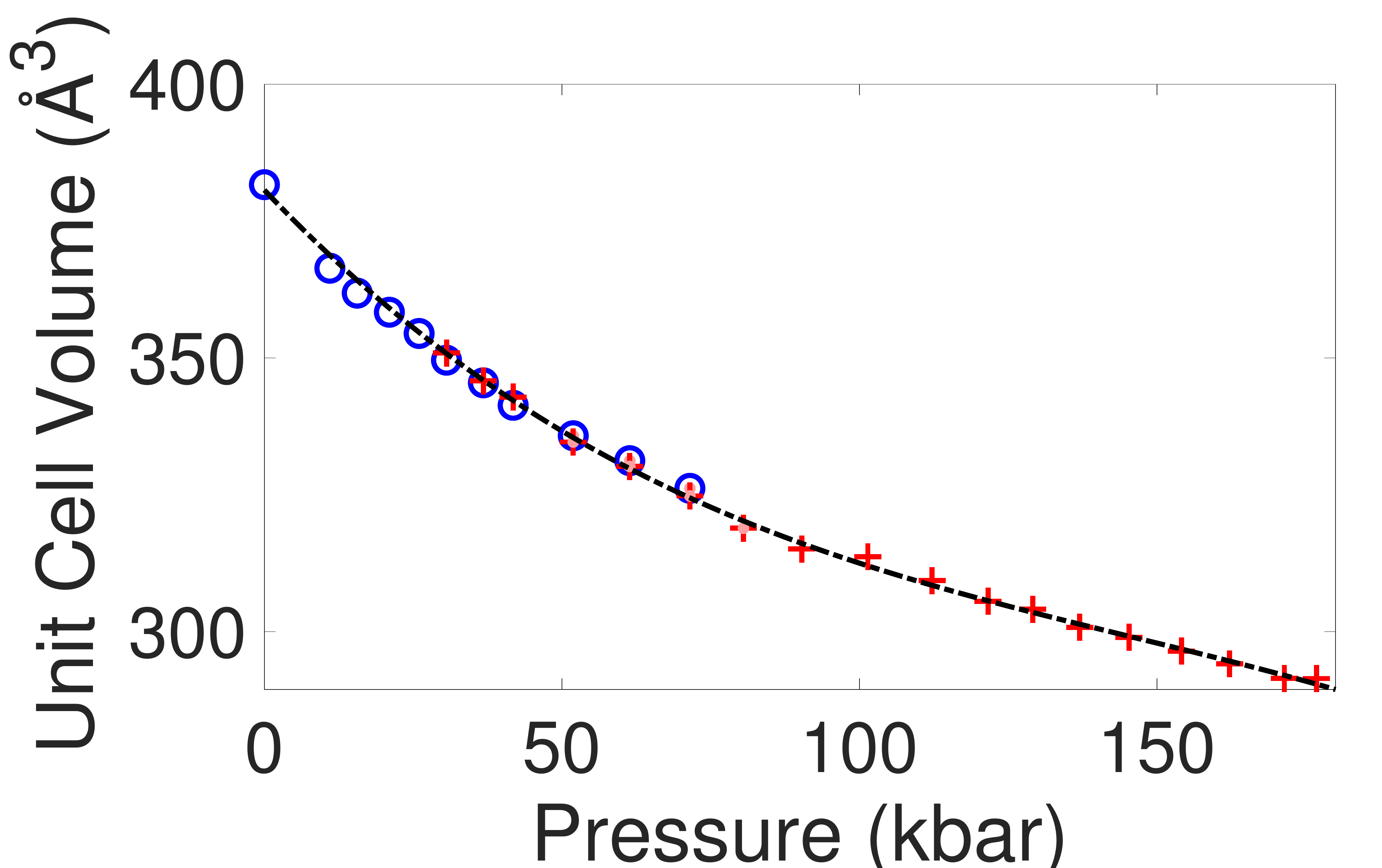}\caption{\label{fig:CrystalStructuresAndParams}Crystal structures and parameters
of the high and low pressure phases. a) - crystal structure of \vpps{}
at 11~kbar - the LP phase and b) - at 177~kbar - the HP-I phase.
The HP-I structure has $\beta$ close to 90\textdegree{} so atoms in each plane
are aligned with their equivalents in neighboring planes. c) and d)
- projections of the same structures along the $b$ axis. e) - phase
fractions of the LP (blue circles) and HP-I (red crosses) phases as
a function of pressure. f) - refined unit cell volumes for each phase.
The LP phase volume is plotted as blue circles and the HP-I as red
crosses. The two phases coexist over a wide pressure range, rather
than an abrupt transition between the two, and the unit cell volume
shows no jumps or sudden transitions.}
\end{figure}

High-pressure powder x-ray diffraction patterns were taken at room
temperature up to 177~kbar. Besides an expected decreasing unit volume
as the sample is pressurized, no changes in the diffraction patterns
from the ambient pressure patterns, and hence the structure described
by Ouvrard et.al. \citep{Ouvrard1985a}, were observed up to 26~kbar.
From 26 to 80~kbar however, a gradual transition to an alternative
high-pressure structure was observed (see Fig. \ref{fig:CrystalStructuresAndParams}).
This new phase (we will denote the ambient and low pressure structural
phase LP and this high-pressure phase HP-I) can be attributed to the
same structure seen in \feps{} at intermediate pressures and designated
HP-I by Haines et.al. \citep{Haines2018b}. The layers of \vpps{}
shift relative to each other in a sliding motion of $\sim a/3$ along
the $a$-axis such that the S atoms become arranged in a hexagonal
close packing layout between the layers, resulting in the monoclinic
unit cell's $\beta$ angle shifting from a value of 107\textdegree{} in the LP
to a value close to 90\textdegree{} (90.13\textdegree{} at 177~ kbar) in the HP-I structure.
In this structure the P atoms are slightly distorted along the $a$-axis
($x$ coordinate value of 0.0074 at 177~kbar) of the unit cell, and
this distortion results in the same C2/m symmetry in the HP-I structure.
In the absence of the distortion of P atoms, HP-I would have a trigonal
symmetry, but certain peak shapes can not be adequately fitted in
refinements made with this space group so we conclude that it remains
monoclinic. As there is no symmetry or even volume change associated
with this LP - HP-I phase transition, it is consistent to observe
it to occur so gradually over a large pressure range, in both \feps{}
and now \vpps{}. Integrated x-ray diffraction patterns, Rietveld
refinements and relevant parameters are shown in the Supplementary
Material, SM at the end of this manuscript.

There are no sudden or discontinuous changes in the cell volume accompanying
this shift; the $c$ lattice spacing changes as its orientation is
altered, but this does not reflect a change in the inter-layer spacing.
The HP-I to HP-II first-order structural transition observed in \feps{}
\citep{Haines2018b} and linked there with the metallization, was
not observed in \vpps{}. In fact, no transitions or distortions of
the HP-I phase were observed up to the maximum pressure measured -
177~kbar. 

\subsection*{Resistivity and insulator-metal transition}

\begin{figure}
\begin{raggedright}
a)
\par\end{raggedright}
\begin{centering}
\includegraphics[width=1\columnwidth]{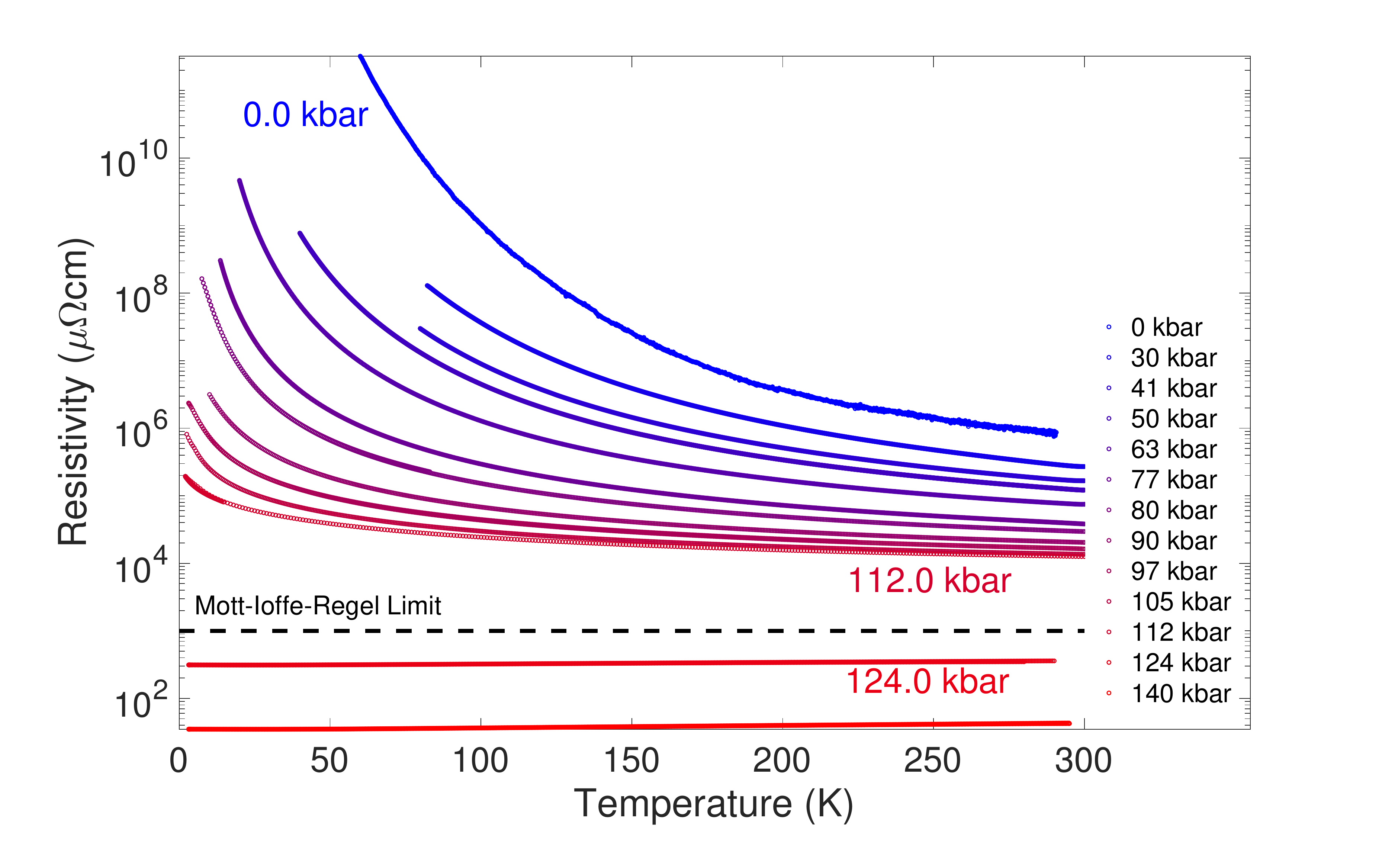}
\par\end{centering}
\begin{raggedright}
b)
\par\end{raggedright}
\centering{}\includegraphics[width=1\columnwidth]{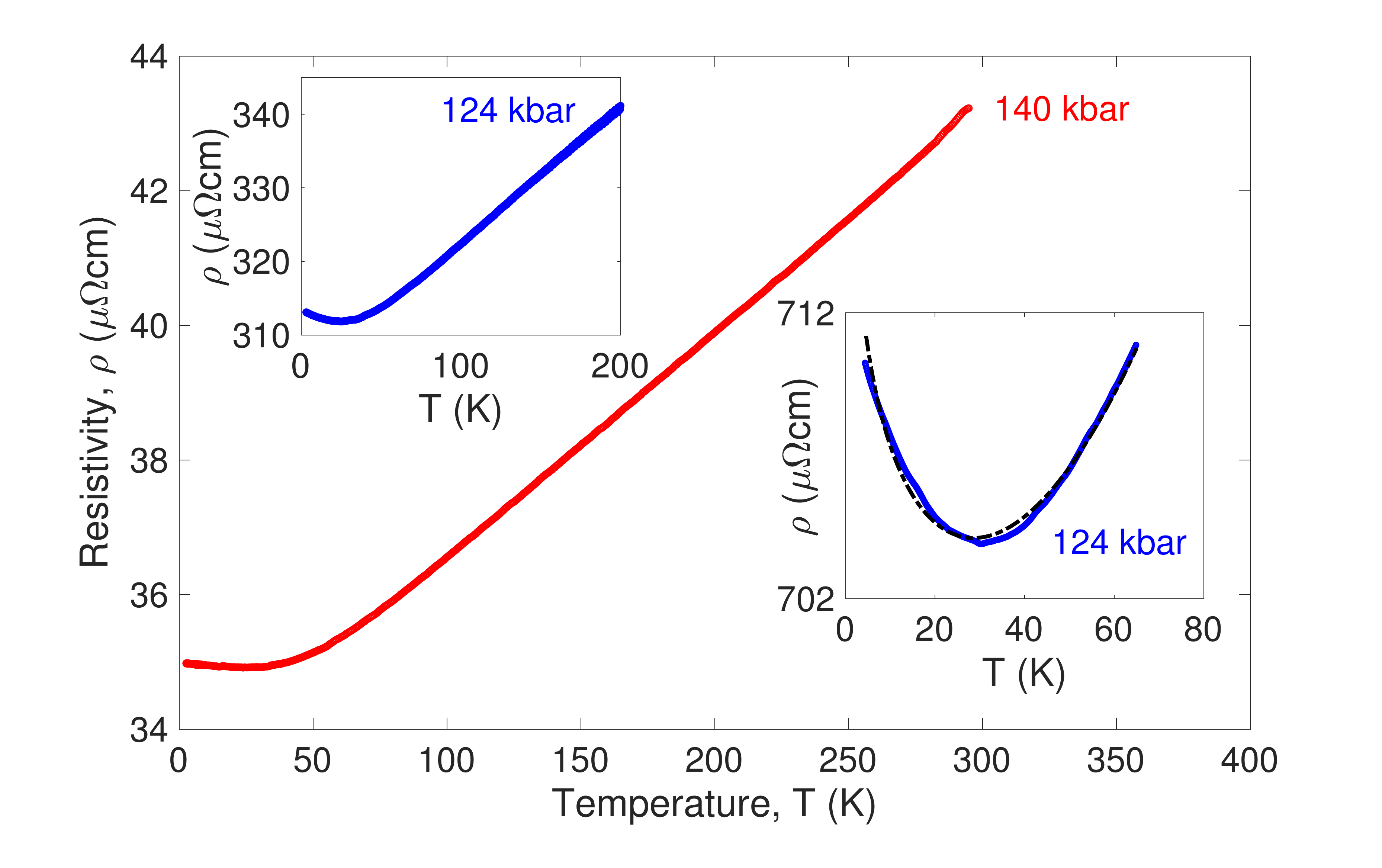}\caption{\label{fig:ResisitivityData}Resistivity data showing the insulator-metal
transition. a) - resistivity plotted against temperature for \vpps{},
for pressures ranging from ambient (blue, topmost) to 140~kbar (red,
bottom). Above 112~kbar the resistivity transitions from an insulating
temperature dependence to a metallic, decreasing with decreasing temperature.
An estimate of the Mott-Ioffe-Regel Limit, as discussed in the text,
is shown as a dotted line, and clearly separates the insulating and
metallic regimes. b) resistivity of \vpps{} at 140~kbar in the metallic
state. Data at 124~kbar, close to the transition, are shown in the
left inset. Right inset shows the low-temperature detail of the data
at 124~kbar with a fit to a Kondo Effect expression as a black dashed
line.}
\end{figure}

\begin{figure}
\centering{}\includegraphics[width=1\columnwidth]{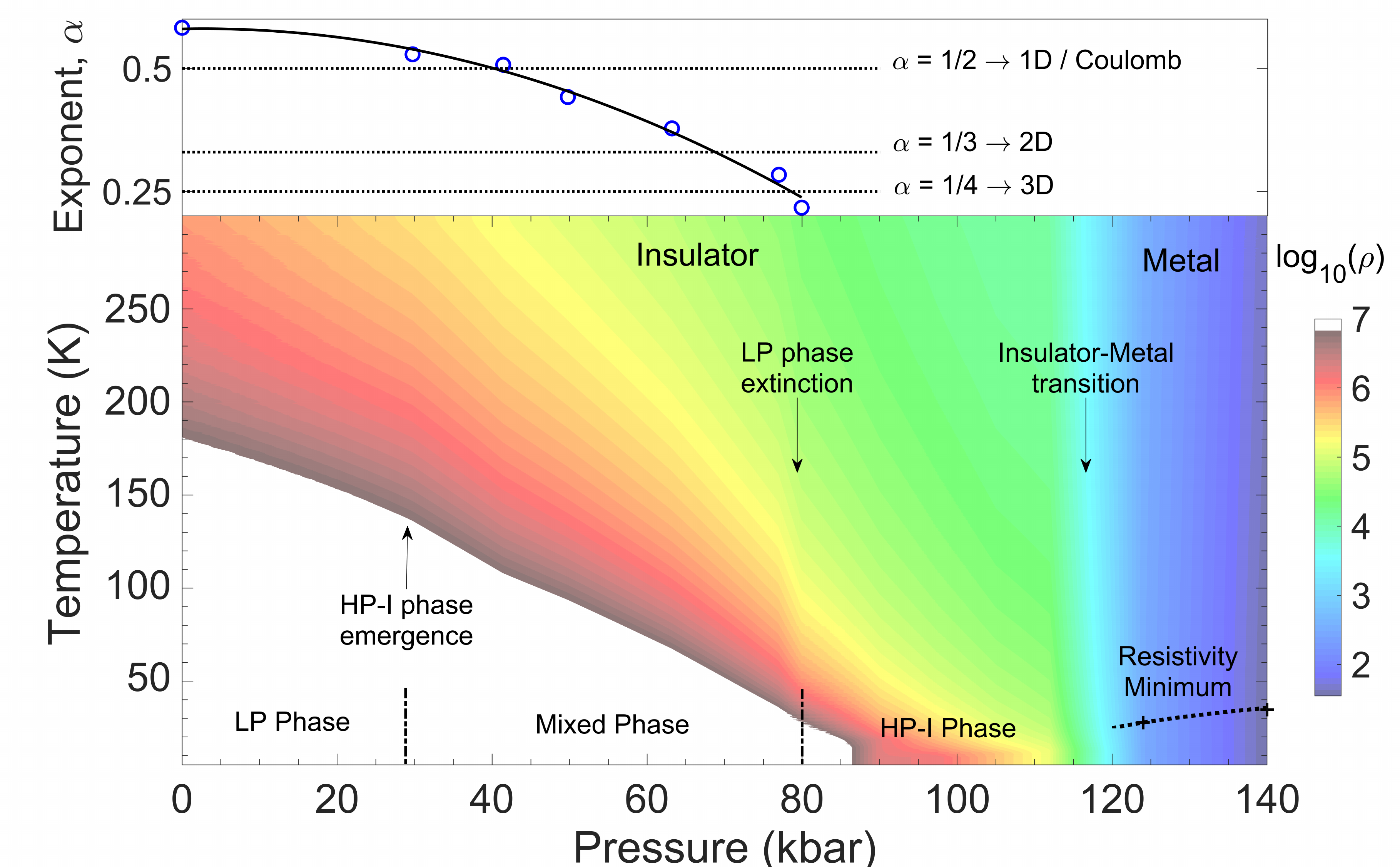}\caption{\label{fig:ResisitivityHeatMap}Resistivity plotted against temperature
and pressure on a logarithmic color scale for \vpps{}. The pressures
corresponding to the beginning and end of the gradual LP - HP-I structural
transition and the insulator-metal transition are marked with arrows.
Upper panel shows pressure dependent values of the exponent $\alpha$
extracted from the variable-range hopping fits described in the text,
with dotted lines to show the gradual increase in effective transport
dimensionality these represent. }
\end{figure}

The temperature dependence of the resistivity $\rho$ of a single
crystal of \vpps{} is shown in Fig. \ref{fig:ResisitivityData}.a
for pressures $p$ ranging from ambient up to 140~kbar. The ambient
pressure resistivity ($8\times10^{5}\mu\Omega$cm) and an energy gap
fitted from an Arrhenius $e^{E_{a}/k_{b}T}$ form (0.2~eV) are consistent
with values previously reported \citep{Ichimura1991} for \vpsevenps{}
and \vps{} - these are substantially lower than all other members
of the \mps{} family. As pressure is increased, the overall magnitude
of the resistivity is dramatically and continuously reduced, and the
curves become shallower, suggesting a reduction of the effective band
gap. Between 112~kbar and 124~kbar the resistivity switches from
an increasing trend with decreasing temperature to a decreasing trend
- the insulator-metal transition. An order-of magnitude estimate of
the Mott-Ioffe-Regel limit \citep{Mott1990a}, following the treatment
of Kurosaki et.al. \citep{Kurosaki2005}, is superimposed and falls
between the insulating and metallic resistivity curves as expected.
Besides the crossover from insulating to metallic behavior, there
appear to be no sudden changes or transitions in the temperature dependence
of the resistivity as pressure is increased - the evolution of the
curves is smooth and continuous. As discussed in the previous section,
there is a structural distortion from the LP to HP-I structure over
the range 26-80~kbar, but no structural changes at pressures above
this - the insulator-metal transition observed in the resistivity
is not accompanied by any structural changes. An isostructural Mott
transition such as this is a very rare phenomenon, particularly in
van-der-Waals materials. Previous examples include specific transition-metal-dichalcogenide
systems \citep{Nayak2014,Nayak2015}, but an equivalent kink in lattice
parameter pressure dependence to that seen in these cases was not
observed here in \vpps{} - there is no signature of the transition
in the structure at all.

The resistivity in the high-pressure metallic state is replotted in
detail in Fig. \ref{fig:ResisitivityData}.b. The resistivity shows
a linear temperature dependence down to around 70~K, and then exhibits
a flattening off and upturn below 40~K. The residual resistance ratio
$R_{300K}/R_{2K}$ is very low at around 1.2, as one would expect
for a highly disordered system like \vpps{}. As the lower inset shows,
the upturn in the low temperature data can be described by the Kondo
effect \citep{Kondo1964}, but alternative forms of localization could
also be responsible for this feature. 

Interestingly, and in contrast to the case of other \mpx{} materials
such as \feps{} \citep{Haines2018b}, the resistivity cannot be well
described by a simple Arrhenius-type insulating temperature dependence
- see supplementary material. The data were found to be best described
by a generalized variable-range-hopping (VRH) \citep{Mott1969,Hill1976,Mott1990a}
expression for highly locally disordered systems $\rho=\rho_{0}Te^{(T_{0}/T)^{\alpha}}$.
The inclusion of a $T^{n}$ prefactor in the exponential resistivity
of an insulator is a common method to describe the thermal dependence
of scatterers in the system - we find a $T$-linear prefactor \citep{Keuls1997}
to best fit the data. The exponent $\alpha$ is given by $\alpha=1/(d+1)$
with $d$ the effective dimensionality of the system; $T_{0}$ is
a characteristic temperature or energy scale of the electron hopping
process. 

Fig. \ref{fig:ResisitivityHeatMap} plots the resistivity against
pressure and temperature, showing the full phase diagram of \vpps{}.
The insulator-metal transition is clearly visible as the point where
resistivity no longer increases with decreasing temperature. The end
point of the LP-HP-I mixed structural phase where the LP phase fraction
goes to zero is accompanied by a visible kink in the resistivity curves
at 80~kbar. The upper panel gives the pressure dependence of the
VRH exponent $\alpha$, with dotted lines showing the values expected
for 1D, 2D and 3D systems of $1/2$, $1/3$ and $1/4$. A strong pressure
dependence of this exponent is observed, with $\alpha$ continuously
decreasing in value as pressure is increased and the insulator-metal
transition approached - the fits lose validity at pressures close
to the transition. The $T_{o}$ characteristic temperature is also
continuously suppressed (see supplementary material) as electron overlap
is increased. As the application of pressure narrows the van-der-Waals
gap between the crystal planes and increases hopping and tunneling
between them, we can reasonably expect a gradual crossover from 2D
to 3D conduction mechanisms. The greater overlap and correlation of
vanadium sites across the crystal planes in HP-I, as well as the important
P$_2$S$_6$-cluster conduction pathways will also bear a role. However
the apparent one dimensional hopping at low pressures is less easily
explained. One explanation could be one-dimensional `chains' of
conduction percolating through the lattice, with the continuous planes
seen in other \mpx{} materials broken up by the vanadium vacancies
and valence mixing. Dimerization of the vanadium-vanadium bonds due
to the vacancies could again contribute to this picture. Another possibility
is that the system is exhibiting so-called Efros-Shklovskii variable-range-hopping
(ES-VRH) \citep{Efros1975,Li2017,Rosenbaum1991} which takes the same
form as the standard Mott VRH but with an exponent $\alpha$ of 0.5,
independent of dimensionality. ES-VRH results from the inclusion of
electron-electron interactions and the development of a Coulomb gap
at the Fermi level below a temperature characteristic of this gap. 

\section*{Discussion}

We have demonstrated a continuous transition from insulating to metallic
states in 2D antiferromagnet \vpps{}. No change in the crystal lattice
was observed in the vicinity of the transition, in contrast to previous
results on \feps{} and \mnps{} where the insulator-metal transition
is accompanied by a dramatic first order structural phase transition:
a collapse of the inter-planar spacing. Mott's original and simplest
explanation of the Mott transition \citep{Mott1990a} involves the
gradual closing of the split metallic bands as the strength of electron
hopping is increased, an eventual touching of the bands causing the
metallization. This mechanism, rather than a structural change, appears
to match the observed behavior in \vpps{}.

\vpps{} does, however, undergo a structural transition or distortion
- over the wide pressure range 26 to 80~kbar a new structural phase
emerges attributed to a sliding motion of the crystal planes. This
brings the $c$ axis to approximately perpendicular to the planes,
and hence the honeycombs of vanadium ions are no longer offset between
planes. This is then consistent with an increase in hopping dimensionality.
The sulfur atoms enter a hexagonal close packed configuration, and
this HP-I structure is very close to possessing a trigonal symmetry
- a slight distortion in the phosphorus positions results in it belonging
to the same monoclinic space group as the LP structure. This transition,
common to all \mpx{} so far measured, occurs well below the metallization
pressure. Accompanying this distortion is a continuous increase in
the effective dimensionality in the variable-range-hopping expression
found to fit the transport data. \vpps{} forms a unique case - unlike
other \mpx{} materials its resistivity follows a variable-range-hopping
rather than an Arrhenius form, and unlike archetypal 3D VRH-metal
transitions \citep{Pollak1991}, such as in doped silicon, the metallization
process cannot be mapped onto a simple scaling relation. In such cases,
the data follow the same functional form, but $T_{0}$ is continuously
suppressed to zero - this is not the case here as the functional form
of the resistivity is constantly and smoothly altered, in addition
to $T_{0}$, as the effective dimensionality of the electronic transport
increases. The evolution of the VRH exponent from 1D to 2D-like can
be potentially understood as originating from ES-VRH hopping, due
to the formation of a Coulomb gap. As pressure and hence and inter-site
hopping is increased, the Coulomb gap is suppressed and Mott VRH hopping
discovered. This is then subsequently tuned to fully 3D hopping and
metallization. The lattice remains 2D throughout, whereas the transport
properties are continuously tuned between regimes, resulting in eventual
metallization. This is a novel mechanism fundamentally in contrast
to previous results, particularly in other members of this material's
family, and the resistivity behavior close to metallization does not
fit any conventional forms and is yet to be explained.

A metallic temperature dependence of the resistivity was observed
at pressures above 112~kbar, with a low-temperature upturn potentially
due to the Kondo Effect. The dilute magnetic impurities of which the
Kondo Effect is a signature we can tentatively attribute to the vanadium
deficiency and disordered valence mixing on the vanadium sites. As
the system is known to be highly locally disordered however, alternative,
more exotic, forms of localization could be responsible for this effect.

As the Mott transition is isostructural, it is likely to be second
order and could potentially be tuned to a quantum critical point.
We can also suspect from magnetotransport data that the metallization
also involves a transition from antiferromagnetic order to paramagnetism,
as is the case in vanadium oxides \citep{Pergament2013}. If this
is found to be the case via further experiments, it would open the
interesting possibility of a spin liquid phase of exotic nature near
the critical point due to the honeycomb lattice, perhaps through a
Kitaev interaction. And, of course, an extremely challenging but exciting
experiment would be to examine the insulator-metal transition in monolayer,
truly two-dimensional, \vps{}. There additionally exists potential
for the formation of a dimerized valence-bond-solid state \citep{Affleck1987}
at high pressure. The spin $3/2$ V$^{2+}$ positioned on a honeycomb
lattice with antiferromagnetic order is a candidate for such a state,
and our observations, paired with the changes in magnetic moment seen
in the iron compounds at metallization \citep{Wang2018}, consistent
with its formation.

\vps{} has, until now, not been studied beyond its basic properties,
while many other members of the \mpx{} family are enjoying wide attention
for their potential in two-dimensional physics and technological applications.
We have demonstrated that this material has many intriguing opportunities
and puzzles for further work, and have demonstrated an isostructural
Mott transition in a new class of 2D van-der-Waals material for the
first time. The insulator-metal transition and the overall transport
mechanisms contrast strongly to the behavior observed in other members
of this family - more work is required to ascertain exactly why. The
lack of an accompanying structural change suggests that the transport
properties of \vpps{} can be much more easily and responsively switched
than in other van-der-Waals materials, whether by chemical doping,
thin film strain or electrostatic gating. This and the material's
small and highly tunable band gap show great promise for future device
applications based on van-der-Waals materials.

\section*{Methods}

Single crystals of \vpps{} were grown via a chemical vapor transport
method in a two-zone tube furnace at temperatures of 600\textdegree{}C and 350\textdegree{}C
for 1 month using 0.1~g of TeCl$_4$ flux for 1~g of reactants.
Prior to the reaction, the quartz tubes used were cleaned and dried,
loaded with V (99.5\%), P (99.99\%) and S (99.98\%) powders under
an argon atmosphere, then evacuated to $5\times10^{-3}$~mbar with
an oil diffusion pump before sealing. The crystals form with a vanadium
deficiency, due to its natural tendency to V$^{3+}$ valence (the
transition metal in \mpx{} is M$^{2+}$ ), so a 20\% excess of vanadium
powder was added to the reactant mixture to attempt to mitigate this.
Additionally, it is worth noting that these reactions will form V$_2$S$_3$
or VS$_2$ at higher temperatures so it is desirable to keep the hot
zone temperature as low as possible for the reaction (Klingen gives
the solid-state reaction temperatures for \mpx{} \citep{Klingen1973})
while allowing sufficient heat for the flux to function. Crystals
were characterized by powder and single-crystal diffraction for phase
purity and by EDX for stoichiometry. The samples used in this study
had stoichiometry of \vpps{}, with an uncertainty of $\pm0.05$ on
the 0.9 vanadium fraction.

The pressure evolution of the crystal structure was found from powder
x-ray diffraction carried out at room temperature on the I15 beamline
at the Diamond Light Source. The powder sample used was ground under
an argon atmosphere (to prevent water uptake) and in liquid nitrogen
to attempt to mitigate the effects of preferred orientation. Helium
was used as the pressure-transmitting medium and the shift in fluorescence
wavelength of ruby spheres placed inside the high pressure region
was used as the pressure calibrant \citep{Mao1986}. An x-ray energy
of 29.2~KeV (\textgreek{l} = 0.4246 \r{A}) was used to collect the diffraction
patterns. A MAR345 2D detector with pixel size 100x100 $\mu$m was
used to record the diffraction patterns with 120~s exposure times
and a 24\textdegree{} rocking of the sample. The data were initially processed
using Dawn \citep{Filik2017} (with a LaB$_6$ calibration), the subsequent
Rietveld refinements were calculated using the GSAS-II software package
\citep{Toby2013} and the structures visualized in VESTA \citep{Momma2011}.
For the structural refinements, a spherical harmonics model for the
observed peak heights was used to take into account the strong (and
pressure-dependent) effects of preferred orientation.

Resistivity measurements were performed on single crystals using a
Keithley 2410 Source Meter with a fixed supplied current of 0.01~$\mu$A
at ambient pressure, and for high pressures using the internal resistance
bridge of the PPMS (Quantum Design) cryostat used for temperature
control. To prepare the samples for these measurements, they were
first mechanically cleaved to expose clean surfaces and a 50~nm layer
of gold was then sputtered onto the surface to form contact pads via
a foil mask. Gold wires were then bonded to these using Dupont 6838
silver epoxy, cured at 180\textdegree{}C for one hour.

A diamond anvil cell \citep{Dunstan1989,Dunstan1989a} with 1~mm
anvil culets and heat-treated Be-Cu gasket was used for the high-pressure
resistivity measurements. Glycerol was used as the pressure medium
and ruby was used to determine the pressure as with the x-ray study.
Estimated pressure uncertainties are $\pm1$~kbar.
\begin{acknowledgments}
This work was carried out with the support of the Diamond Light Source
and we acknowledge the provision of beamtime at I15 under proposal
number NT21368. The authors would like to thank P.A.C. Brown, S.E.
Dutton, I.Hwang D. Jarvis and Y. Noda for their generous help and
discussions. We would also like to acknowledge support from Jesus
College of the University of Cambridge, IHT KAZATOMPROM and the CHT
Uzbekistan programme. The work was carried out with financial support
from the Ministry of Education and Science of the Russian Federation
in the framework of Increase Competitiveness Program of NUST MISiS
(\textnumero{} \textcyr{\CYRK}2-2017-024). This work was supported
by the Institute for Basic Science (IBS) in Korea (Grant No. IBS-R009-G1).
\end{acknowledgments}

\bibliographystyle{apsrev4-1}

\begin{thebibliography}{53}
\bibitem [{ {Ajayan}\ \emph {et~al.}(2016) {Ajayan},
   {Kim},\ and\  {Banerjee}}]{Ajayan2016}%
  \BibitemOpen
  \bibfield  {author} {\bibinfo {author} { {P.}~
  {Ajayan}}, \bibinfo {author} { {P.}~ {Kim}}, \ and\
  \bibinfo {author} { {K.}~ {Banerjee}},\ }\href
  {\doibase 10.1063/pt.3.3297} {\bibfield  {journal} {\bibinfo  {journal}
  {Physics Today}\ }\textbf {\bibinfo {volume} {69}},\ \bibinfo {pages} {38}
  (\bibinfo {year} {2016})}\BibitemShut {NoStop}%
\bibitem [{ {Park}(2016)}]{Park2016}%
  \BibitemOpen
  \bibfield  {author} {\bibinfo {author} { {J.-G.}\ 
  {Park}},\ }\href {\doibase 10.1088/0953-8984/28/30/301001} {\bibfield
  {journal} {\bibinfo  {journal} {J. Phys.: Condens. Matter}\ }\textbf
  {\bibinfo {volume} {28}},\ \bibinfo {pages} {301001} (\bibinfo {year}
  {2016})}\BibitemShut {NoStop}%
\bibitem [{ {Samarth}(2017)}]{Samarth2017}%
  \BibitemOpen
  \bibfield  {author} {\bibinfo {author} { {N.}~
  {Samarth}},\ }\href {\doibase 10.1038/546216a} {\bibfield  {journal}
  {\bibinfo  {journal} {Nature}\ }\textbf {\bibinfo {volume} {546}},\ \bibinfo
  {pages} {216} (\bibinfo {year} {2017})}\BibitemShut {NoStop}%
\bibitem [{ {Zhou}\ \emph {et~al.}(2016) {Zhou},
   {Lu},  {Zu},\ and\ 
  {Gao}}]{Zhou2016}%
  \BibitemOpen
  \bibfield  {author} {\bibinfo {author} { {Y.}~
  {Zhou}}, \bibinfo {author} { {H.}~ {Lu}}, \bibinfo
  {author} { {X.}~ {Zu}}, \ and\ \bibinfo {author}
  { {F.}~ {Gao}},\ }\href {\doibase 10.1038/srep19407}
  {\bibfield  {journal} {\bibinfo  {journal} {Scientific Reports}\ }\textbf
  {\bibinfo {volume} {6}} (\bibinfo {year} {2016}),\
  10.1038/srep19407}\BibitemShut {NoStop}%
\bibitem [{ {Burch}\ \emph {et~al.}(2018) {Burch},
   {Mandrus},\ and\  {Park}}]{Burch2018}%
  \BibitemOpen
  \bibfield  {author} {\bibinfo {author} { {K.~S.}\ 
  {Burch}}, \bibinfo {author} { {D.}~ {Mandrus}}, \
  and\ \bibinfo {author} { {J.-G.}\  {Park}},\ }\href
  {\doibase 10.1038/s41586-018-0631-z} {\bibfield  {journal} {\bibinfo
  {journal} {Nature}\ }\textbf {\bibinfo {volume} {563}},\ \bibinfo {pages}
  {47} (\bibinfo {year} {2018})}\BibitemShut {NoStop}%
\bibitem [{ {Klingen}\ \emph {et~al.}(1968)
  {Klingen},  {Eulenberger},\ and\ 
  {Hahn}}]{Klingen1968}%
  \BibitemOpen
  \bibfield  {author} {\bibinfo {author} { {W.}~
  {Klingen}}, \bibinfo {author} { {G.}~
  {Eulenberger}}, \ and\ \bibinfo {author} { {H.}~
  {Hahn}},\ }\href {\doibase 10.1007/bf00606219} {\bibfield  {journal}
  {\bibinfo  {journal} {Die Naturwissenschaften}\ }\textbf {\bibinfo {volume}
  {55}},\ \bibinfo {pages} {229} (\bibinfo {year} {1968})}\BibitemShut
  {NoStop}%
\bibitem [{ {Klingen}\ \emph {et~al.}(1970)
  {Klingen},  {Eulenberger},\ and\ 
  {Hahn}}]{Klingen1970}%
  \BibitemOpen
  \bibfield  {author} {\bibinfo {author} { {W.}~
  {Klingen}}, \bibinfo {author} { {G.}~
  {Eulenberger}}, \ and\ \bibinfo {author} { {H.}~
  {Hahn}},\ }\href {\doibase 10.1007/BF00590690} {\bibfield  {journal}
  {\bibinfo  {journal} {Naturwissenschaften}\ }\textbf {\bibinfo {volume}
  {57}},\ \bibinfo {pages} {88} (\bibinfo {year} {1970})}\BibitemShut {NoStop}%
\bibitem [{ {Klingen}\ \emph {et~al.}(1973)
  {Klingen},  {Ott},\ and\  {Hahn}}]{Klingen1973}%
  \BibitemOpen
  \bibfield  {author} {\bibinfo {author} { {W.}~
  {Klingen}}, \bibinfo {author} { {R.}~ {Ott}}, \ and\
  \bibinfo {author} { {H.}~ {Hahn}},\ }\href {\doibase
  10.1002/zaac.19733960305} {\bibfield  {journal} {\bibinfo  {journal}
  {Zeitschrift f\"ur anorganische und allgemeine Chemie}\ }\textbf {\bibinfo
  {volume} {396}},\ \bibinfo {pages} {271} (\bibinfo {year}
  {1973})}\BibitemShut {NoStop}%
\bibitem [{ {Grasso}\ and\ 
  {Silipigni}(2002)}]{Grasso2002}%
  \BibitemOpen
  \bibfield  {author} {\bibinfo {author} { {V.}~
  {Grasso}}\ and\ \bibinfo {author} { {L.}~
  {Silipigni}},\ }\href {http://adsabs.harvard.edu/abs/2002NCimR..25f...1G}
  {\bibfield  {journal} {\bibinfo  {journal} {Rivista Del Nuovo Cimento}\
  }\textbf {\bibinfo {volume} {25}},\ \bibinfo {pages} {2002} (\bibinfo {year}
  {2002})}\BibitemShut {NoStop}%
\bibitem [{ {Kurosawa}\ \emph {et~al.}(1983)
  {Kurosawa},  {Saito},\ and\ 
  {Yamaguchi}}]{Kurosawa1983}%
  \BibitemOpen
  \bibfield  {author} {\bibinfo {author} { {K.}~
  {Kurosawa}}, \bibinfo {author} { {S.}~ {Saito}}, \
  and\ \bibinfo {author} { {Y.}~ {Yamaguchi}},\ }\href
  {http://jpsj.ipap.jp/link?JPSJ/52/3919/} {\bibfield  {journal} {\bibinfo
  {journal} {Journal of the Physical Society of Japan}\ }\textbf {\bibinfo
  {volume} {52}},\ \bibinfo {pages} {3919} (\bibinfo {year}
  {1983})}\BibitemShut {NoStop}%
\bibitem [{ {Okuda}\ \emph {et~al.}(1986) {Okuda},
   {Kurosawa},  {Saito},  {Honda},
   {Yu},\ and\  {Date}}]{Okuda1986}%
  \BibitemOpen
  \bibfield  {author} {\bibinfo {author} { {K.}~
  {Okuda}}, \bibinfo {author} { {K.}~ {Kurosawa}},
  \bibinfo {author} { {S.}~ {Saito}}, \bibinfo
  {author} { {M.}~ {Honda}}, \bibinfo {author}
  { {Z.}~ {Yu}}, \ and\ \bibinfo {author}
  { {M.}~ {Date}},\ }\href {\doibase
  10.1143/jpsj.55.4456} {\bibfield  {journal} {\bibinfo  {journal} {J. Phys.
  Soc. Jpn.}\ }\textbf {\bibinfo {volume} {55}},\ \bibinfo {pages} {4456}
  (\bibinfo {year} {1986})}\BibitemShut {NoStop}%
\bibitem [{ {Wildes}\ \emph {et~al.}(1994) {Wildes},
   {Kennedy},\ and\  {Hicks}}]{Wildes1994}%
  \BibitemOpen
  \bibfield  {author} {\bibinfo {author} { {A.}~
  {Wildes}}, \bibinfo {author} { {S.}~ {Kennedy}}, \
  and\ \bibinfo {author} { {T.}~ {Hicks}},\ }\href
  {\doibase 10.1088/0953-8984/6/24/002} {\bibfield  {journal} {\bibinfo
  {journal} {J. Phys.: Condens. Matter}\ }\textbf {\bibinfo {volume} {6}},\
  \bibinfo {pages} {L335} (\bibinfo {year} {1994})}\BibitemShut {NoStop}%
\bibitem [{ {Wildes}\ \emph {et~al.}(1998) {Wildes},
   {Roessli},  {Lebech},\ and\ 
  {Godfrey}}]{Wildes1998a}%
  \BibitemOpen
  \bibfield  {author} {\bibinfo {author} { {A.}~
  {Wildes}}, \bibinfo {author} { {B.}~ {Roessli}},
  \bibinfo {author} { {B.}~ {Lebech}}, \ and\ \bibinfo
  {author} { {K.}~ {Godfrey}},\ }\href {\doibase
  10.1088/0953-8984/10/28/020} {\bibfield  {journal} {\bibinfo  {journal} {J.
  Phys.: Condens. Matter}\ }\textbf {\bibinfo {volume} {10}},\ \bibinfo {pages}
  {6417} (\bibinfo {year} {1998})}\BibitemShut {NoStop}%
\bibitem [{ {Rule}\ \emph {et~al.}(2007) {Rule},
   {McIntyre},  {Kennedy},\ and\ 
  {Hicks}}]{Rule2007}%
  \BibitemOpen
  \bibfield  {author} {\bibinfo {author} { {K.}~
  {Rule}}, \bibinfo {author} { {G.}~ {McIntyre}},
  \bibinfo {author} { {S.}~ {Kennedy}}, \ and\
  \bibinfo {author} { {T.}~ {Hicks}},\ }\href
  {http://prb.aps.org/abstract/PRB/v76/i13/e134402} {\bibfield  {journal}
  {\bibinfo  {journal} {Phys. Rev. B}\ }\textbf {\bibinfo {volume} {76}},\
  \bibinfo {pages} {134402} (\bibinfo {year} {2007})}\BibitemShut {NoStop}%
\bibitem [{ {Wildes}\ \emph {et~al.}(2007) {Wildes},
   {R{\o}nnow},  {Roessli},  {Harris},\
  and\  {Godfrey}}]{Wildes2007}%
  \BibitemOpen
  \bibfield  {author} {\bibinfo {author} { {A.}~
  {Wildes}}, \bibinfo {author} { {H.}~ {R{\o}nnow}},
  \bibinfo {author} { {B.}~ {Roessli}}, \bibinfo
  {author} { {M.}~ {Harris}}, \ and\ \bibinfo {author}
  { {K.}~ {Godfrey}},\ }\href {\doibase
  10.1016/j.jmmm.2006.10.347} {\bibfield  {journal} {\bibinfo  {journal}
  {Journal of Magnetism and Magnetic Materials}\ }\textbf {\bibinfo {volume}
  {310}},\ \bibinfo {pages} {1221} (\bibinfo {year} {2007})}\BibitemShut
  {NoStop}%
\bibitem [{ {Wildes}\ \emph {et~al.}(2012) {Wildes},
   {Rule},  {Bewley},  {Enderle},\ and\
   {Hicks}}]{Wildes2012}%
  \BibitemOpen
  \bibfield  {author} {\bibinfo {author} { {A.}~
  {Wildes}}, \bibinfo {author} { {K.}~ {Rule}},
  \bibinfo {author} { {R.}~ {Bewley}}, \bibinfo
  {author} { {M.}~ {Enderle}}, \ and\ \bibinfo
  {author} { {T.}~ {Hicks}},\ }\href {\doibase
  10.1088/0953-8984/24/41/416004} {\bibfield  {journal} {\bibinfo  {journal}
  {J. Phys.: Condens. Matter}\ }\textbf {\bibinfo {volume} {24}},\ \bibinfo
  {pages} {416004} (\bibinfo {year} {2012})}\BibitemShut {NoStop}%
\bibitem [{ {Wildes}\ \emph {et~al.}(2015) {Wildes},
   {Simonet},  {Ressouche},  {McIntyre},
   {Avdeev},  {Suard},  {Kimber},
   {Lan{\c{c}}on},  {Pepe}, 
  {Moubaraki},\ and\  {et~al.}}]{Wildes2015}%
  \BibitemOpen
  \bibfield  {author} {\bibinfo {author} { {A.}~
  {Wildes}}, \bibinfo {author} { {V.}~ {Simonet}},
  \bibinfo {author} { {E.}~ {Ressouche}}, \bibinfo
  {author} { {G.}~ {McIntyre}}, \bibinfo {author}
  { {M.}~ {Avdeev}}, \bibinfo {author} {
  {E.}~ {Suard}}, \bibinfo {author} {
  {S.}~ {Kimber}}, \bibinfo {author} {
  {D.}~ {Lan{\c{c}}on}}, \bibinfo {author} {
  {G.}~ {Pepe}}, \bibinfo {author} { {B.}~
  {Moubaraki}}, \ and\ \bibinfo {author} { {et~al.}},\ }\href
  {\doibase 10.1103/physrevb.92.224408} {\bibfield  {journal} {\bibinfo
  {journal} {Physical Review B}\ }\textbf {\bibinfo {volume} {92}} (\bibinfo
  {year} {2015}),\ 10.1103/physrevb.92.224408}\BibitemShut {NoStop}%
\bibitem [{ {Wildes}\ \emph {et~al.}(2017) {Wildes},
   {Simonet},  {Ressouche},  {Ballou},\
  and\  {{McIntyre}}}]{Wildes2017}%
  \BibitemOpen
  \bibfield  {author} {\bibinfo {author} { {A.}~
  {Wildes}}, \bibinfo {author} { {V.}~ {Simonet}},
  \bibinfo {author} { {E.}~ {Ressouche}}, \bibinfo
  {author} { {R.}~ {Ballou}}, \ and\ \bibinfo {author}
  { {G.}~ {{McIntyre}}},\ }\href {\doibase
  10.1088/1361-648X/aa8a43} {\bibfield  {journal} {\bibinfo  {journal} {J.
  Phys.: Condens. Matter}\ }\textbf {\bibinfo {volume} {29}},\ \bibinfo {pages}
  {455801} (\bibinfo {year} {2017})}\BibitemShut {NoStop}%
\bibitem [{ {Lan{\c{c}}on}\ \emph {et~al.}(2016)
  {Lan{\c{c}}on},  {Walker},  {Ressouche},
   {Ouladdiaf},  {Rule},  {McIntyre},
   {Hicks},  {R{\o}nnow},\ and\ 
  {Wildes}}]{Lancon_2016}%
  \BibitemOpen
  \bibfield  {author} {\bibinfo {author} { {D.}~
  {Lan{\c{c}}on}}, \bibinfo {author} { {H.}~
  {Walker}}, \bibinfo {author} { {E.}~ {Ressouche}},
  \bibinfo {author} { {B.}~ {Ouladdiaf}}, \bibinfo
  {author} { {K.}~ {Rule}}, \bibinfo {author}
  { {G.}~ {McIntyre}}, \bibinfo {author}
  { {T.}~ {Hicks}}, \bibinfo {author} {
  {H.}~ {R{\o}nnow}}, \ and\ \bibinfo {author} {
  {A.}~ {Wildes}},\ }\href {\doibase 10.1103/physrevb.94.214407}
  {\bibfield  {journal} {\bibinfo  {journal} {Physical Review B}\ }\textbf
  {\bibinfo {volume} {94}} (\bibinfo {year} {2016}),\
  10.1103/physrevb.94.214407}\BibitemShut {NoStop}%
\bibitem [{ {Brec}\ \emph {et~al.}(1979) {Brec},
   {Schleich},  {Ouvrard},  {Louisy},\
  and\  {Rouxel}}]{Brec1979}%
  \BibitemOpen
  \bibfield  {author} {\bibinfo {author} { {R.}~
  {Brec}}, \bibinfo {author} { {D.~M.}\  {Schleich}},
  \bibinfo {author} { {G.}~ {Ouvrard}}, \bibinfo
  {author} { {A.}~ {Louisy}}, \ and\ \bibinfo {author}
  { {J.}~ {Rouxel}},\ }\href
  {http://pubs.acs.org/doi/abs/10.1021/ic50197a018} {\bibfield  {journal}
  {\bibinfo  {journal} {Inorganic Chemistry}\ }\textbf {\bibinfo {volume}
  {18}},\ \bibinfo {pages} {1814} (\bibinfo {year} {1979})}\BibitemShut
  {NoStop}%
\bibitem [{ {Ouvrard}\ \emph
  {et~al.}(1985{\natexlab{a}}) {Ouvrard},  {Brec},\
  and\  {Rouxel}}]{Ouvrard1985}%
  \BibitemOpen
  \bibfield  {author} {\bibinfo {author} { {G.}~
  {Ouvrard}}, \bibinfo {author} { {R.}~ {Brec}}, \
  and\ \bibinfo {author} { {J.}~ {Rouxel}},\ }\href
  {\doibase 10.1016/0025-5408(85)90092-3} {\bibfield  {journal} {\bibinfo
  {journal} {Materials Research Bulletin}\ }\textbf {\bibinfo {volume} {20}},\
  \bibinfo {pages} {1181} (\bibinfo {year} {1985}{\natexlab{a}})}\BibitemShut
  {NoStop}%
\bibitem [{ {Ouvrard}\ \emph
  {et~al.}(1985{\natexlab{b}}) {Ouvrard},  {Fr\'eour},
   {Brec},\ and\  {Rouxel}}]{Ouvrard1985a}%
  \BibitemOpen
  \bibfield  {author} {\bibinfo {author} { {G.}~
  {Ouvrard}}, \bibinfo {author} { {R.}~ {Fr\'eour}},
  \bibinfo {author} { {R.}~ {Brec}}, \ and\ \bibinfo
  {author} { {J.}~ {Rouxel}},\ }\href {\doibase
  10.1016/0025-5408(85)90204-1} {\bibfield  {journal} {\bibinfo  {journal}
  {Materials Research Bulletin}\ }\textbf {\bibinfo {volume} {20}},\ \bibinfo
  {pages} {1053} (\bibinfo {year} {1985}{\natexlab{b}})}\BibitemShut {NoStop}%
\bibitem [{ {Brec}(1986)}]{Brec1986}%
  \BibitemOpen
  \bibfield  {author} {\bibinfo {author} { {R.}~
  {Brec}},\ }\href {\doibase 10.1016/0167-2738(86)90055-x} {\bibfield
  {journal} {\bibinfo  {journal} {Solid State Ionics}\ }\textbf {\bibinfo
  {volume} {22}},\ \bibinfo {pages} {3} (\bibinfo {year} {1986})}\BibitemShut
  {NoStop}%
\bibitem [{ {Lee}\ \emph {et~al.}(2016) {Lee},
   {Lee},  {Ryoo},  {Kang},
   {Kim},  {Kim},  {Park}, 
  {Park},\ and\  {Cheong}}]{Lee_2016}%
  \BibitemOpen
  \bibfield  {author} {\bibinfo {author} { {J.}~
  {Lee}}, \bibinfo {author} { {S.}~ {Lee}}, \bibinfo
  {author} { {J.}~ {Ryoo}}, \bibinfo {author}
  { {S.}~ {Kang}}, \bibinfo {author} {
  {T.}~ {Kim}}, \bibinfo {author} { {P.}~
  {Kim}}, \bibinfo {author} { {C.}~ {Park}}, \bibinfo
  {author} { {J.}~ {Park}}, \ and\ \bibinfo {author}
  { {H.}~ {Cheong}},\ }\href {\doibase
  10.1021/acs.nanolett.6b03052} {\bibfield  {journal} {\bibinfo  {journal}
  {Nano Letters}\ }\textbf {\bibinfo {volume} {16}},\ \bibinfo {pages} {7433}
  (\bibinfo {year} {2016})}\BibitemShut {NoStop}%
\bibitem [{ {Kuo}\ \emph {et~al.}(2016) {Kuo},
   {Neumann},  {Balamurugan},  {Park},
   {Kang},  {Shiu},  {Kang},
   {Hong},  {Han},  {Noh},\ and\
   {Park}}]{Kuo2016}%
  \BibitemOpen
  \bibfield  {author} {\bibinfo {author} { {C.}~
  {Kuo}}, \bibinfo {author} { {M.}~ {Neumann}},
  \bibinfo {author} { {K.}~ {Balamurugan}}, \bibinfo
  {author} { {H.}~ {Park}}, \bibinfo {author}
  { {S.}~ {Kang}}, \bibinfo {author} {
  {H.}~ {Shiu}}, \bibinfo {author} { {J.}~
  {Kang}}, \bibinfo {author} { {B.}~ {Hong}}, \bibinfo
  {author} { {M.}~ {Han}}, \bibinfo {author}
  { {T.}~ {Noh}}, \ and\ \bibinfo {author}
  { {J.-G.}\  {Park}},\ }\href {\doibase
  10.1038/srep20904} {\bibfield  {journal} {\bibinfo  {journal} {Scientific
  Reports}\ }\textbf {\bibinfo {volume} {6}},\ \bibinfo {pages} {20904}
  (\bibinfo {year} {2016})}\BibitemShut {NoStop}%
\bibitem [{ {Haines}\ \emph {et~al.}(2018) {Haines},
   {Coak},  {Wildes},  {Lampronti},
   {Liu},  {Nahai-Williamson}, 
  {Hamidov},  {Daisenberger},\ and\ 
  {Saxena}}]{Haines2018b}%
  \BibitemOpen
  \bibfield  {author} {\bibinfo {author} { {C.}~
  {Haines}}, \bibinfo {author} { {M.}~ {Coak}},
  \bibinfo {author} { {A.}~ {Wildes}}, \bibinfo
  {author} { {G.}~ {Lampronti}}, \bibinfo {author}
  { {C.}~ {Liu}}, \bibinfo {author} {
  {P.}~ {Nahai-Williamson}}, \bibinfo {author} {
  {H.}~ {Hamidov}}, \bibinfo {author} {
  {D.}~ {Daisenberger}}, \ and\ \bibinfo {author} {
  {S.}~ {Saxena}},\ }\href {\doibase
  10.1103/physrevlett.121.266801} {\bibfield  {journal} {\bibinfo  {journal}
  {Physical Review Letters}\ }\textbf {\bibinfo {volume} {121}} (\bibinfo
  {year} {2018}),\ 10.1103/physrevlett.121.266801}\BibitemShut {NoStop}%
\bibitem [{ {Wang}\ \emph {et~al.}(2016) {Wang},
   {Zhou},  {Wen},  {Zhou},
   {Li},  {Han},  {Xiao}, 
  {Chow},  {Sun},  {Pravica}, 
  {Cornelius},  {Yang},\ and\  {Zhao}}]{Wang2016a}%
  \BibitemOpen
  \bibfield  {author} {\bibinfo {author} { {Y.}~
  {Wang}}, \bibinfo {author} { {Z.}~ {Zhou}}, \bibinfo
  {author} { {T.}~ {Wen}}, \bibinfo {author}
  { {Y.}~ {Zhou}}, \bibinfo {author} {
  {N.}~ {Li}}, \bibinfo {author} { {F.}~
  {Han}}, \bibinfo {author} { {Y.}~ {Xiao}}, \bibinfo
  {author} { {P.}~ {Chow}}, \bibinfo {author}
  { {J.}~ {Sun}}, \bibinfo {author} {
  {M.}~ {Pravica}}, \bibinfo {author} { {A.~L.}\
   {Cornelius}}, \bibinfo {author} { {W.}~
  {Yang}}, \ and\ \bibinfo {author} { {Y.}~ {Zhao}},\
  }\href {\doibase 10.1021/jacs.6b10225} {\bibfield  {journal} {\bibinfo
  {journal} {Journal of the American Chemical Society}\ }\textbf {\bibinfo
  {volume} {138}},\ \bibinfo {pages} {15751} (\bibinfo {year}
  {2016})}\BibitemShut {NoStop}%
\bibitem [{ {Tsurubayashi}\ \emph {et~al.}(2018)
  {Tsurubayashi},  {Kodama},  {Kano}, 
  {Ishigaki},  {Uwatoko},  {Watanabe}, 
  {Takase},\ and\  {Takano}}]{Tsurubayashi2018}%
  \BibitemOpen
  \bibfield  {author} {\bibinfo {author} { {M.}~
  {Tsurubayashi}}, \bibinfo {author} { {K.}~
  {Kodama}}, \bibinfo {author} { {M.}~ {Kano}},
  \bibinfo {author} { {K.}~ {Ishigaki}}, \bibinfo
  {author} { {Y.}~ {Uwatoko}}, \bibinfo {author}
  { {T.}~ {Watanabe}}, \bibinfo {author}
  { {K.}~ {Takase}}, \ and\ \bibinfo {author}
  { {Y.}~ {Takano}},\ }\href {\doibase
  10.1063/1.5043121} {\bibfield  {journal} {\bibinfo  {journal} {{AIP}
  Advances}\ }\textbf {\bibinfo {volume} {8}},\ \bibinfo {pages} {101307}
  (\bibinfo {year} {2018})}\BibitemShut {NoStop}%
\bibitem [{ {Wang}\ \emph {et~al.}(2018) {Wang},
   {Ying},  {Zhou},  {Sun},
   {Wen},  {Zhou},  {Li}, 
  {Zhang},  {Han},  {Xiao},  {Chow},
   {Yang},  {Struzhkin},  {Zhao},\ and\
   {kwang Mao}}]{Wang2018}%
  \BibitemOpen
  \bibfield  {author} {\bibinfo {author} { {Y.}~
  {Wang}}, \bibinfo {author} { {J.}~ {Ying}}, \bibinfo
  {author} { {Z.}~ {Zhou}}, \bibinfo {author}
  { {J.}~ {Sun}}, \bibinfo {author} {
  {T.}~ {Wen}}, \bibinfo {author} { {Y.}~
  {Zhou}}, \bibinfo {author} { {N.}~ {Li}}, \bibinfo
  {author} { {Q.}~ {Zhang}}, \bibinfo {author}
  { {F.}~ {Han}}, \bibinfo {author} {
  {Y.}~ {Xiao}}, \bibinfo {author} { {P.}~
  {Chow}}, \bibinfo {author} { {W.}~ {Yang}}, \bibinfo
  {author} { {V.~V.}\  {Struzhkin}}, \bibinfo {author}
  { {Y.}~ {Zhao}}, \ and\ \bibinfo {author}
  { {H.}~ {kwang Mao}},\ }\href {\doibase
  10.1038/s41467-018-04326-1} {\bibfield  {journal} {\bibinfo  {journal}
  {Nature Communications}\ }\textbf {\bibinfo {volume} {9}} (\bibinfo {year}
  {2018}),\ 10.1038/s41467-018-04326-1}\BibitemShut {NoStop}%
\bibitem [{ {Brec}\ and\  {Rouxel}(1980)}]{Brec1980}%
  \BibitemOpen
  \bibfield  {author} {\bibinfo {author} { {R.}~
  {Brec}}\ and\ \bibinfo {author} { {J.}~ {Rouxel}},\
  } {\emph {\bibinfo {booktitle} {New Ways to Save Energy}}}\
  (\bibinfo  {publisher} {Springer},\ \bibinfo {year} {1980})\ pp.\ \bibinfo
  {pages} {620--630}\BibitemShut {NoStop}%
\bibitem [{ {Ichimura}\ and\ 
  {Sano}(1991)}]{Ichimura1991}%
  \BibitemOpen
  \bibfield  {author} {\bibinfo {author} { {K.}~
  {Ichimura}}\ and\ \bibinfo {author} { {M.}~
  {Sano}},\ }\href {\doibase 10.1016/0379-6779(91)91804-j} {\bibfield
  {journal} {\bibinfo  {journal} {Synthetic Metals}\ }\textbf {\bibinfo
  {volume} {45}},\ \bibinfo {pages} {203} (\bibinfo {year} {1991})}\BibitemShut
  {NoStop}%
\bibitem [{ {Chittari}\ \emph {et~al.}(2016)
  {Chittari},  {Park},  {Lee},  {Han},
   {{MacDonald}},  {Hwang},\ and\ 
  {Jung}}]{Chittari2016}%
  \BibitemOpen
  \bibfield  {author} {\bibinfo {author} { {B.}~
  {Chittari}}, \bibinfo {author} { {Y.}~ {Park}},
  \bibinfo {author} { {D.}~ {Lee}}, \bibinfo {author}
  { {M.}~ {Han}}, \bibinfo {author} {
  {A.}~ {{MacDonald}}}, \bibinfo {author} {
  {E.}~ {Hwang}}, \ and\ \bibinfo {author} {
  {J.}~ {Jung}},\ }\href {\doibase 10.1103/PhysRevB.94.184428}
  {\bibfield  {journal} {\bibinfo  {journal} {Phys.Rev.B}\ }\textbf {\bibinfo
  {volume} {94}},\ \bibinfo {pages} {184428} (\bibinfo {year}
  {2016})}\BibitemShut {NoStop}%
\bibitem [{ {Mott}(1990)}]{Mott1990a}%
  \BibitemOpen
  \bibfield  {author} {\bibinfo {author} { {N.}~
  {Mott}},\ } {\emph {\bibinfo {title} {Metal-Insulator
  Transitions}}}\ (\bibinfo  {publisher} {Taylor and Francis, London},\
  \bibinfo {year} {1990})\BibitemShut {NoStop}%
\bibitem [{ {Kurosaki}\ \emph {et~al.}(2005)
  {Kurosaki},  {Shimizu},  {Miyagawa}, 
  {Kanoda},\ and\  {Saito}}]{Kurosaki2005}%
  \BibitemOpen
  \bibfield  {author} {\bibinfo {author} { {Y.}~
  {Kurosaki}}, \bibinfo {author} { {Y.}~ {Shimizu}},
  \bibinfo {author} { {K.}~ {Miyagawa}}, \bibinfo
  {author} { {K.}~ {Kanoda}}, \ and\ \bibinfo {author}
  { {G.}~ {Saito}},\ }\href {\doibase
  10.1103/physrevlett.95.177001} {\bibfield  {journal} {\bibinfo  {journal}
  {Physical Review Letters}\ }\textbf {\bibinfo {volume} {95}} (\bibinfo {year}
  {2005}),\ 10.1103/physrevlett.95.177001}\BibitemShut {NoStop}%
\bibitem [{ {Nayak}\ \emph {et~al.}(2014) {Nayak},
   {Bhattacharyya},  {Zhu},  {Liu},
   {Wu},  {Pandey},  {Jin},
   {Singh},  {Akinwande},\ and\ 
  {Lin}}]{Nayak2014}%
  \BibitemOpen
  \bibfield  {author} {\bibinfo {author} { {A.}~
  {Nayak}}, \bibinfo {author} { {S.}~
  {Bhattacharyya}}, \bibinfo {author} { {J.}~ {Zhu}},
  \bibinfo {author} { {J.}~ {Liu}}, \bibinfo {author}
  { {X.}~ {Wu}}, \bibinfo {author} {
  {T.}~ {Pandey}}, \bibinfo {author} {
  {C.}~ {Jin}}, \bibinfo {author} { {A.}~
  {Singh}}, \bibinfo {author} { {D.}~ {Akinwande}}, \
  and\ \bibinfo {author} { {J.-F.}\  {Lin}},\ }\href
  {\doibase 10.1038/ncomms4731} {\bibfield  {journal} {\bibinfo  {journal}
  {Nature Communications}\ }\textbf {\bibinfo {volume} {5}} (\bibinfo {year}
  {2014}),\ 10.1038/ncomms4731}\BibitemShut {NoStop}%
\bibitem [{ {Nayak}\ \emph {et~al.}(2015) {Nayak},
   {Yuan},  {Cao},  {Liu}, 
  {Wu},  {Moran},  {Li},  {Akinwande},
   {Jin},\ and\  {Lin}}]{Nayak2015}%
  \BibitemOpen
  \bibfield  {author} {\bibinfo {author} { {A.}~
  {Nayak}}, \bibinfo {author} { {Z.}~ {Yuan}},
  \bibinfo {author} { {B.}~ {Cao}}, \bibinfo {author}
  { {J.}~ {Liu}}, \bibinfo {author} {
  {J.}~ {Wu}}, \bibinfo {author} { {S.}~
  {Moran}}, \bibinfo {author} { {T.}~ {Li}}, \bibinfo
  {author} { {D.}~ {Akinwande}}, \bibinfo {author}
  { {C.}~ {Jin}}, \ and\ \bibinfo {author}
  { {J.-F.}\  {Lin}},\ }\href {\doibase
  10.1021/acsnano.5b03295} {\bibfield  {journal} {\bibinfo  {journal} {{ACS}
  Nano}\ }\textbf {\bibinfo {volume} {9}},\ \bibinfo {pages} {9117} (\bibinfo
  {year} {2015})}\BibitemShut {NoStop}%
\bibitem [{ {Kondo}(1964)}]{Kondo1964}%
  \BibitemOpen
  \bibfield  {author} {\bibinfo {author} { {J.}~
  {Kondo}},\ }\href {\doibase 10.1143/ptp.32.37} {\bibfield  {journal}
  {\bibinfo  {journal} {Progress of Theoretical Physics}\ }\textbf {\bibinfo
  {volume} {32}},\ \bibinfo {pages} {37} (\bibinfo {year} {1964})}\BibitemShut
  {NoStop}%
\bibitem [{ {Mott}(1969)}]{Mott1969}%
  \BibitemOpen
  \bibfield  {author} {\bibinfo {author} { {N.}~
  {Mott}},\ }in\ \href {\doibase 10.1016/b978-0-08-015543-2.50005-x} {\emph
  {\bibinfo {booktitle} {Festk\"orper Probleme {IX}}}}\ (\bibinfo  {publisher}
  {Elsevier},\ \bibinfo {year} {1969})\ pp.\ \bibinfo {pages}
  {22--45}\BibitemShut {NoStop}%
\bibitem [{ {Hill}(1976)}]{Hill1976}%
  \BibitemOpen
  \bibfield  {author} {\bibinfo {author} { {R.}~
  {Hill}},\ }\href {\doibase 10.1002/pssa.2210340223} {\bibfield  {journal}
  {\bibinfo  {journal} {Physica Status Solidi (a)}\ }\textbf {\bibinfo {volume}
  {34}},\ \bibinfo {pages} {601} (\bibinfo {year} {1976})}\BibitemShut
  {NoStop}%
\bibitem [{ {Keuls}\ \emph {et~al.}(1997) {Keuls},
   {Hu},  {Jiang},\ and\ 
  {Dahm}}]{Keuls1997}%
  \BibitemOpen
  \bibfield  {author} {\bibinfo {author} { {F.~V.}\ 
  {Keuls}}, \bibinfo {author} { {X.}~ {Hu}}, \bibinfo
  {author} { {H.}~ {Jiang}}, \ and\ \bibinfo {author}
  { {A.}~ {Dahm}},\ }\href {\doibase
  10.1103/physrevb.56.1161} {\bibfield  {journal} {\bibinfo  {journal}
  {Physical Review B}\ }\textbf {\bibinfo {volume} {56}},\ \bibinfo {pages}
  {1161} (\bibinfo {year} {1997})}\BibitemShut {NoStop}%
\bibitem [{ {Efros}\ and\ 
  {Shklovskii}(1975)}]{Efros1975}%
  \BibitemOpen
  \bibfield  {author} {\bibinfo {author} { {A.}~
  {Efros}}\ and\ \bibinfo {author} { {B.}~
  {Shklovskii}},\ }\href {\doibase 10.1088/0022-3719/8/4/003} {\bibfield
  {journal} {\bibinfo  {journal} {Journal of Physics C: Solid State Physics}\
  }\textbf {\bibinfo {volume} {8}},\ \bibinfo {pages} {L49} (\bibinfo {year}
  {1975})}\BibitemShut {NoStop}%
\bibitem [{ {Li}\ \emph {et~al.}(2017) {Li},
   {Peng},  {Zhang},  {Li},
   {Zeng},  {Luo},  {Zhan},
   {Meng},  {Zhou},\ and\ 
  {Wu}}]{Li2017}%
  \BibitemOpen
  \bibfield  {author} {\bibinfo {author} { {Z.}~
  {Li}}, \bibinfo {author} { {L.}~ {Peng}}, \bibinfo
  {author} { {J.}~ {Zhang}}, \bibinfo {author}
  { {J.}~ {Li}}, \bibinfo {author} {
  {Y.}~ {Zeng}}, \bibinfo {author} { {Y.}~
  {Luo}}, \bibinfo {author} { {Z.}~ {Zhan}}, \bibinfo
  {author} { {L.}~ {Meng}}, \bibinfo {author}
  { {M.}~ {Zhou}}, \ and\ \bibinfo {author}
  { {W.}~ {Wu}},\ }\href {\doibase
  10.1088/1361-6641/aa5390} {\bibfield  {journal} {\bibinfo  {journal}
  {Semiconductor Science and Technology}\ }\textbf {\bibinfo {volume} {32}},\
  \bibinfo {pages} {035010} (\bibinfo {year} {2017})}\BibitemShut {NoStop}%
\bibitem [{ {Rosenbaum}(1991)}]{Rosenbaum1991}%
  \BibitemOpen
  \bibfield  {author} {\bibinfo {author} { {R.}~
  {Rosenbaum}},\ }\href {\doibase 10.1103/physrevb.44.3599} {\bibfield
  {journal} {\bibinfo  {journal} {Physical Review B}\ }\textbf {\bibinfo
  {volume} {44}},\ \bibinfo {pages} {3599} (\bibinfo {year}
  {1991})}\BibitemShut {NoStop}%
\bibitem [{ {Pollak}\ and\ 
  {Shklovskii}(1991)}]{Pollak1991}%
  \BibitemOpen
  \bibfield  {author} {\bibinfo {author} { {M.}~
  {Pollak}}\ and\ \bibinfo {author} { {B.}~
  {Shklovskii}},\ }
  {\emph {\bibinfo {title} {Hopping transport in solids}}}\ (\bibinfo
  {publisher} {Elsevier},\ \bibinfo {year} {1991})\BibitemShut {NoStop}%
\bibitem [{ {Pergament}\ \emph {et~al.}(2013)
  {Pergament},  {Stefanovich},  {Kuldin},\ and\
   {Velichko}}]{Pergament2013}%
  \BibitemOpen
  \bibfield  {author} {\bibinfo {author} { {A.}~
  {Pergament}}, \bibinfo {author} { {G.}~
  {Stefanovich}}, \bibinfo {author} { {N.}~ {Kuldin}},
  \ and\ \bibinfo {author} { {A.}~ {Velichko}},\
  }\href {\doibase 10.1155/2013/960627} {\bibfield  {journal} {\bibinfo
  {journal} {{ISRN} Condensed Matter Physics}\ }\textbf {\bibinfo {volume}
  {2013}},\ \bibinfo {pages} {1} (\bibinfo {year} {2013})}\BibitemShut
  {NoStop}%
\bibitem [{ {Affleck}\ \emph {et~al.}(1987)
  {Affleck},  {Kennedy},  {Lieb},\ and\ 
  {Tasaki}}]{Affleck1987}%
  \BibitemOpen
  \bibfield  {author} {\bibinfo {author} { {I.}~
  {Affleck}}, \bibinfo {author} { {T.}~ {Kennedy}},
  \bibinfo {author} { {E.}~ {Lieb}}, \ and\ \bibinfo
  {author} { {H.}~ {Tasaki}},\ }
  {\bibfield  {journal} {\bibinfo  {journal} {Phys. Rev. Lett}\ } (\bibinfo
  {year} {1987})}\BibitemShut {NoStop}%
\bibitem [{ {Mao}\ \emph {et~al.}(1986) {Mao},
   {Xu},\ and\  {Bell}}]{Mao1986}%
  \BibitemOpen
  \bibfield  {author} {\bibinfo {author} { {H.}~
  {Mao}}, \bibinfo {author} { {J.}~ {Xu}}, \ and\
  \bibinfo {author} { {P.}~ {Bell}},\ }\href {\doibase
  10.1029/jb091ib05p04673} {\bibfield  {journal} {\bibinfo  {journal} {Journal
  of Geophysical Research}\ }\textbf {\bibinfo {volume} {91}},\ \bibinfo
  {pages} {4673} (\bibinfo {year} {1986})}\BibitemShut {NoStop}%
\bibitem [{ {Filik}\ \emph {et~al.}(2017) {Filik},
   {Ashton},  {Chang},  {Chater},
   {Day},  {Drakopoulos},  {Gerring},
   {Hart},  {Magdysyuk},  {Michalik},
   {Smith},  {Tang},  {Terrill},
   {Wharmby},\ and\  {Wilhelm}}]{Filik2017}%
  \BibitemOpen
  \bibfield  {author} {\bibinfo {author} { {J.}~
  {Filik}}, \bibinfo {author} { {A.}~ {Ashton}},
  \bibinfo {author} { {P.}~ {Chang}}, \bibinfo
  {author} { {P.}~ {Chater}}, \bibinfo {author}
  { {S.}~ {Day}}, \bibinfo {author} {
  {M.}~ {Drakopoulos}}, \bibinfo {author} {
  {M.}~ {Gerring}}, \bibinfo {author} {
  {M.}~ {Hart}}, \bibinfo {author} { {O.}~
  {Magdysyuk}}, \bibinfo {author} { {S.}~ {Michalik}},
  \bibinfo {author} { {A.}~ {Smith}}, \bibinfo
  {author} { {C.}~ {Tang}}, \bibinfo {author}
  { {N.}~ {Terrill}}, \bibinfo {author} {
  {M.}~ {Wharmby}}, \ and\ \bibinfo {author} {
  {H.}~ {Wilhelm}},\ }\href {\doibase 10.1107/s1600576717004708}
  {\bibfield  {journal} {\bibinfo  {journal} {Journal of Applied
  Crystallography}\ }\textbf {\bibinfo {volume} {50}},\ \bibinfo {pages} {959}
  (\bibinfo {year} {2017})}\BibitemShut {NoStop}%
\bibitem [{ {Toby}\ and\  {Dreele}(2013)}]{Toby2013}%
  \BibitemOpen
  \bibfield  {author} {\bibinfo {author} { {B.}~
  {Toby}}\ and\ \bibinfo {author} { {R.~V.}\ 
  {Dreele}},\ }\href {\doibase 10.1107/s0021889813003531} {\bibfield  {journal}
  {\bibinfo  {journal} {Journal of Applied Crystallography}\ }\textbf {\bibinfo
  {volume} {46}},\ \bibinfo {pages} {544} (\bibinfo {year} {2013})}\BibitemShut
  {NoStop}%
\bibitem [{ {Momma}\ and\ 
  {Izumi}(2011)}]{Momma2011}%
  \BibitemOpen
  \bibfield  {author} {\bibinfo {author} { {K.}~
  {Momma}}\ and\ \bibinfo {author} { {F.}~ {Izumi}},\
  }\href {\doibase 10.1107/s0021889811038970} {\bibfield  {journal} {\bibinfo
  {journal} {J Appl Cryst}\ }\textbf {\bibinfo {volume} {44}},\ \bibinfo
  {pages} {1272} (\bibinfo {year} {2011})}\BibitemShut {NoStop}%
\bibitem [{ {Dunstan}\ and\ 
  {Spain}(1989)}]{Dunstan1989}%
  \BibitemOpen
  \bibfield  {author} {\bibinfo {author} { {D.}~
  {Dunstan}}\ and\ \bibinfo {author} { {I.}~
  {Spain}},\ }\href {\doibase 10.1088/0022-3735/22/11/004} {\bibfield
  {journal} {\bibinfo  {journal} {J. Phys. E: Sci. Instrum.}\ }\textbf
  {\bibinfo {volume} {22}},\ \bibinfo {pages} {913} (\bibinfo {year}
  {1989})}\BibitemShut {NoStop}%
\bibitem [{ {Spain}\ and\ 
  {Dunstan}(1989)}]{Dunstan1989a}%
  \BibitemOpen
  \bibfield  {author} {\bibinfo {author} { {I.}~
  {Spain}}\ and\ \bibinfo {author} { {D.}~
  {Dunstan}},\ }\href {\doibase 10.1088/0022-3735/22/11/005} {\bibfield
  {journal} {\bibinfo  {journal} {J. Phys. E: Sci. Instrum.}\ }\textbf
  {\bibinfo {volume} {22}},\ \bibinfo {pages} {923} (\bibinfo {year}
  {1989})}\BibitemShut {NoStop}%
\bibitem [{ {Mao}\ \emph {et~al.}(1988) {Mao},
   {Hemley},  {Wu},  {Jephcoat},
   {Finger},  {Zha},\ and\ 
  {Bassett}}]{Mao1988}%
  \BibitemOpen
  \bibfield  {author} {\bibinfo {author} { {H.}~
  {Mao}}, \bibinfo {author} { {R.}~ {Hemley}},
  \bibinfo {author} { {Y.}~ {Wu}}, \bibinfo {author}
  { {A.}~ {Jephcoat}}, \bibinfo {author}
  { {L.}~ {Finger}}, \bibinfo {author} {
  {C.}~ {Zha}}, \ and\ \bibinfo {author} {
  {W.}~ {Bassett}},\ }\href {\doibase 10.1103/physrevlett.60.2649}
  {\bibfield  {journal} {\bibinfo  {journal} {Physical Review Letters}\
  }\textbf {\bibinfo {volume} {60}},\ \bibinfo {pages} {2649} (\bibinfo {year}
  {1988})}\BibitemShut {NoStop}
\end{thebibliography}

\pagebreak{}

\onecolumngrid

\section*{Supplementary Material}

\begin{figure}[H]
\begin{centering}
\includegraphics[width=0.8\columnwidth]{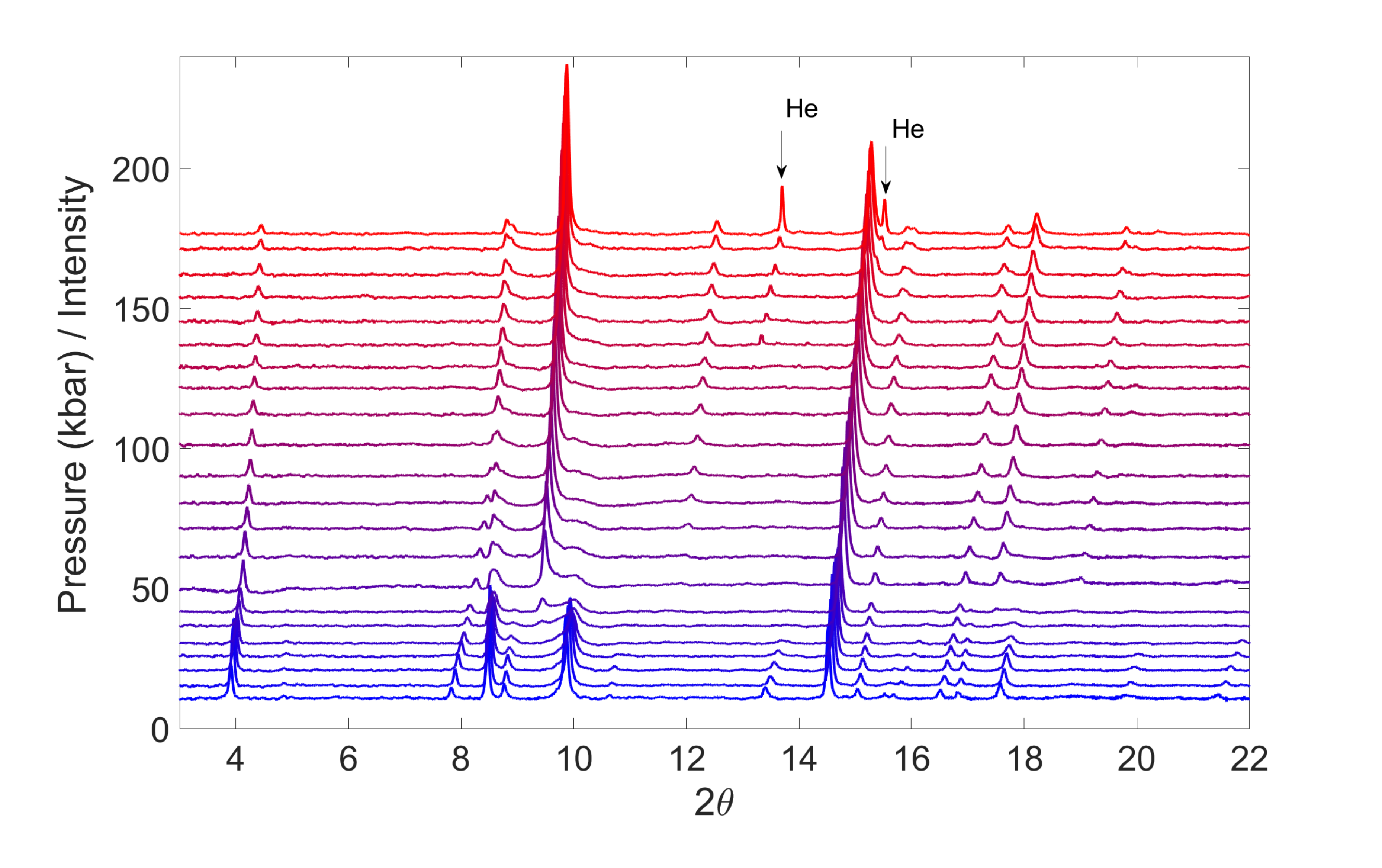}
\par\end{centering}
\centering{}\caption{\label{fig:XRayPatterns}Integrated powder diffraction patterns for
\vpps{} at room temperature. A diffuse background has been subtracted
and the y-axis offset of each curve has been set to the value of its
pressure in kbar.}
\end{figure}

\begin{figure}[H]
\begin{centering}
\includegraphics[width=0.8\columnwidth]{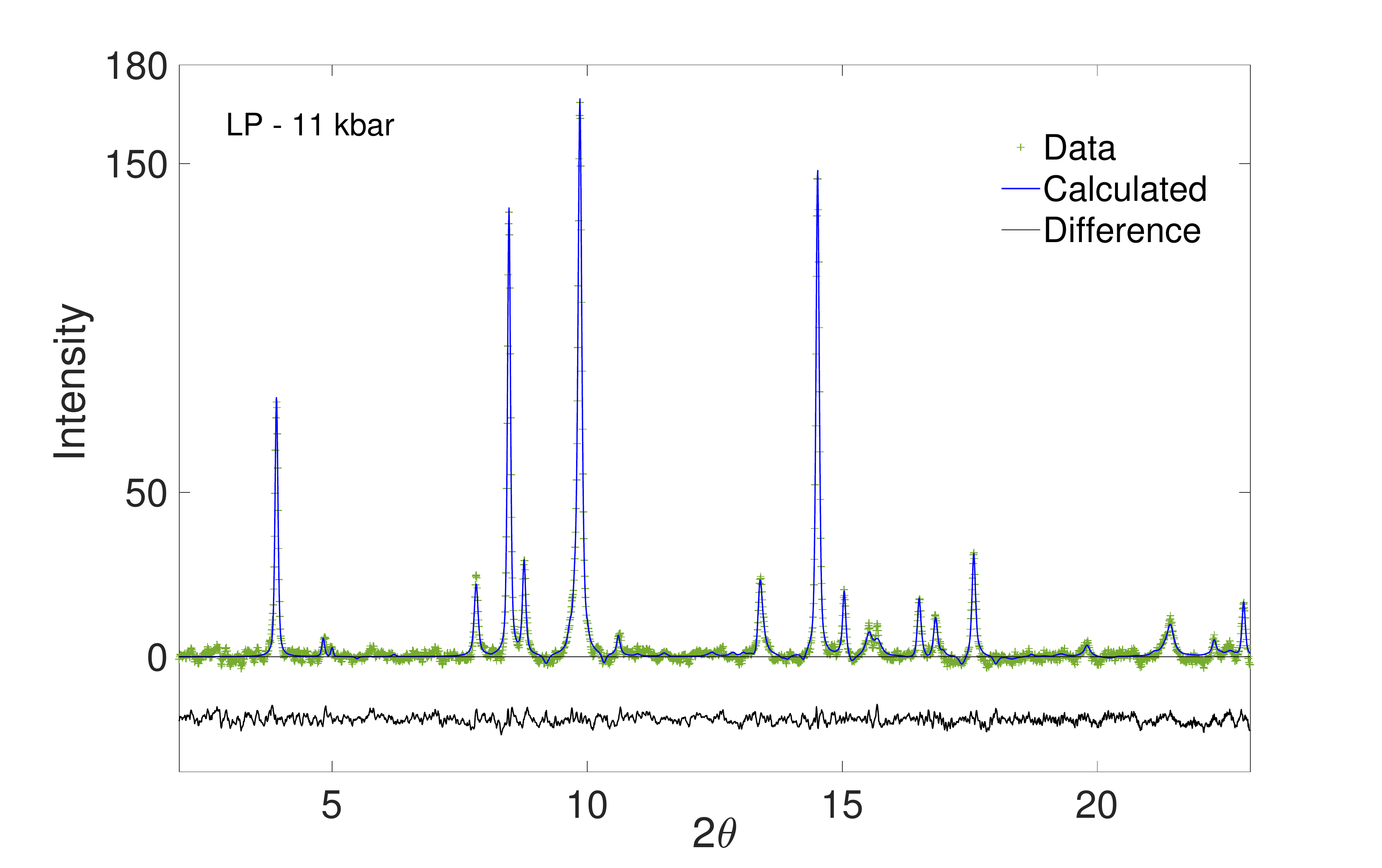}
\par\end{centering}
\centering{}\caption{\label{fig:RefinementPlot-LP}Rietveld refinement of the integrated
powder x-ray patterns, after subtraction of a diffuse background,
for \vpps{} at 11~kbar (the lowest pressure measured), in the LP
phase.}
\end{figure}

\begin{figure}
\centering{}\includegraphics[width=0.8\columnwidth]{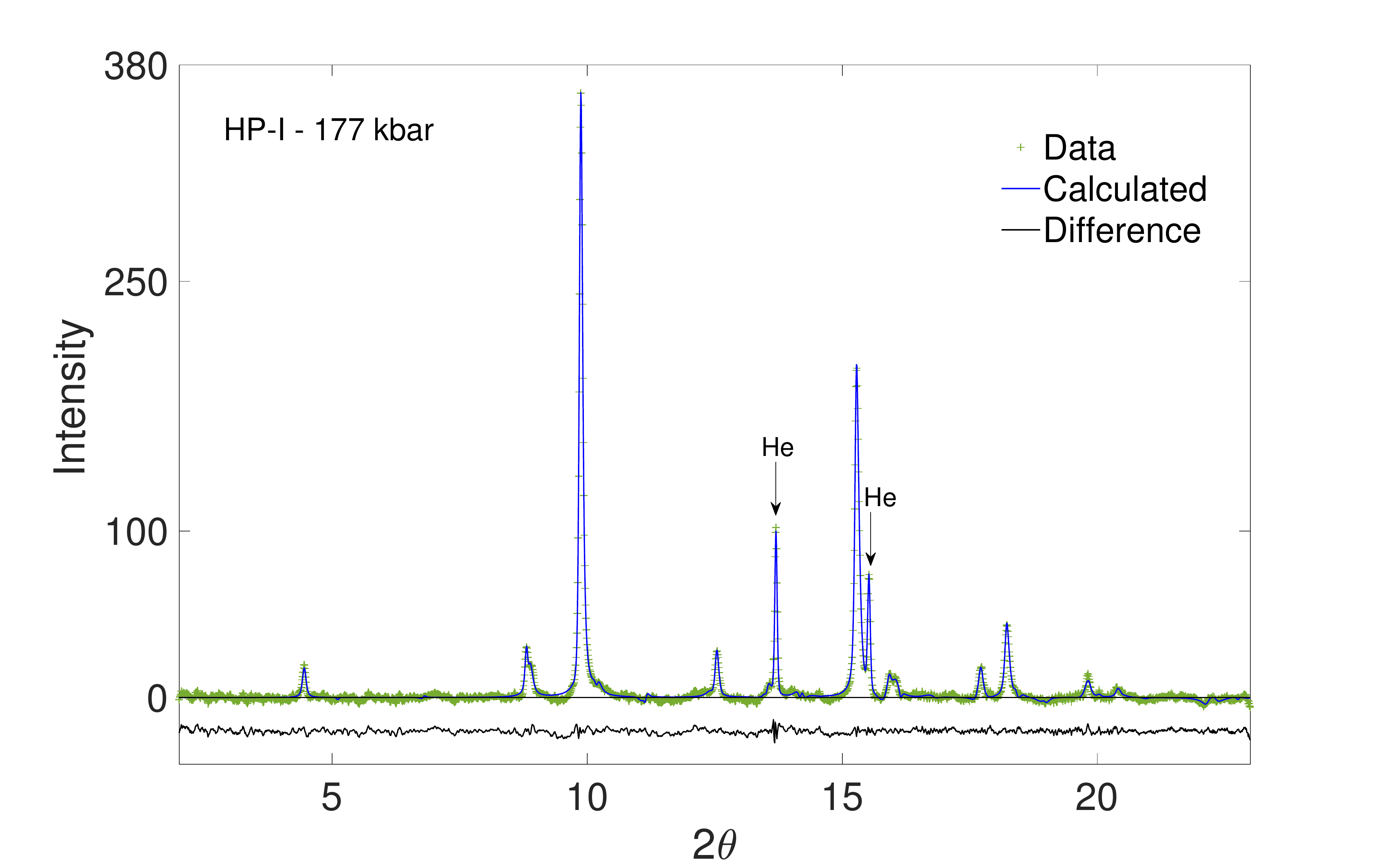}\caption{\label{fig:RefinementPlot_HPI} Rietveld refinement of the integrated
powder x-ray patterns, after subtraction of a diffuse background,
for \vpps{} at 177~kbar, in the HP-I phase. The peaks marked `He'
in are due to the solid helium pressure medium and were included in
the refinement as an additional phase or impurity. These appear at
150~kbar - the freezing point of helium at room temperature, and
the change of their positions with pressure were found to match the
data of Mao et.al. \citep{Mao1988}.}
\end{figure}

\begin{table*}
\begin{centering}
\begin{tabular}{>{\centering}m{0.14\textwidth}>{\centering}m{0.14\textwidth}>{\centering}m{0.14\textwidth}>{\centering}m{0.14\textwidth}>{\centering}m{0.16\textwidth}>{\centering}m{0.16\textwidth}}
LP (11~kbar) & C2/m & $R$ = 7\% & $wR$ = 9\% & $\chi^{2}$ = 3.01 & GOF = 4\%\tabularnewline
\addlinespace[0.0075\textheight]
$a$ = 5.8436(15)~\r{A} & $b$ = 10.0876(8)~\r{A} & $c$ = 6.5237(18)~\r{A} & \textgreek{b} = 107.098(5)~\textdegree{} & V = 367.56(6)~\r{A}$^{3}$ & \textgreek{r} = 3.060(1)~g.cm$^{-3}$\tabularnewline\addlinespace[0.0075\textheight]
\midrule 
 & $x$ & $y$ & $z$ & Occ & $U_{iso}$\tabularnewline
\midrule
V(4g) & 0 & 0.331(1) & 0 & 0.86(4) & 0.079(4)\tabularnewline
P(4i) & 0.043(4) & 0 & 0.122(3) & 1 & 0.091(6)\tabularnewline
S(4i) & 0.758(2) & 0 & 0.249(4) & 1 & 0.071(6)\tabularnewline
S(8j) & 0.260(2) & 0.1738(7) & 0.234(2) & 1 & 0.070(3)\tabularnewline
\bottomrule
\end{tabular}
\par\end{centering}
\vspace{0.8cm}

\begin{centering}
\begin{tabular}{>{\centering}m{0.14\textwidth}>{\centering}m{0.14\textwidth}>{\centering}m{0.14\textwidth}>{\centering}m{0.14\textwidth}>{\centering}m{0.16\textwidth}>{\centering}m{0.16\textwidth}}
HP-I (177~kbar) & C2/m & $R$ = 7\% & $wR$ = 10\% & $\chi^{2}$ = 3.56 & GOF = 4\%\tabularnewline
\addlinespace[0.0075\textheight]
$a$ = 5.5469(3)~\r{A} & $b$ = 9.5892(6)~\r{A} & $c$ = 5.4788(9)~\r{A} & \textgreek{b} = 90.136(8)~\textdegree{} & V = 291.42(7)~\r{A}$^{3}$ & \textgreek{r} = 3.813(1)~g.cm$^{-3}$\tabularnewline\addlinespace[0.0075\textheight]
\midrule
 & $x$ & $y$ & $z$ & Occ & $U_{iso}$\tabularnewline
\midrule
V(4g) & 0 & 0.3431(4) & 0 & 0.86(4) & 0.090(3)\tabularnewline
P(4i) & 0.0076(7) & 0 & 0.8492(20) & 1 & 0.061(3)\tabularnewline
S(4i) & 0.3257(11) & 0 & 0.7214(7) & 1 & 0.025(4)\tabularnewline
S(8j) & 0.8441(3) & 0.1649(3) & 0.7510(5) & 1 & 0.039(3)\tabularnewline
\bottomrule
\end{tabular}
\par\end{centering}
\caption{\label{tab:refinementParamsTable}Refinement parameters for the LP
low-pressure phase at 11~kbar and the HP-I high-pressure phase at
177~kbar.}
\end{table*}

\begin{figure}
\begin{centering}
\includegraphics[width=0.5\columnwidth]{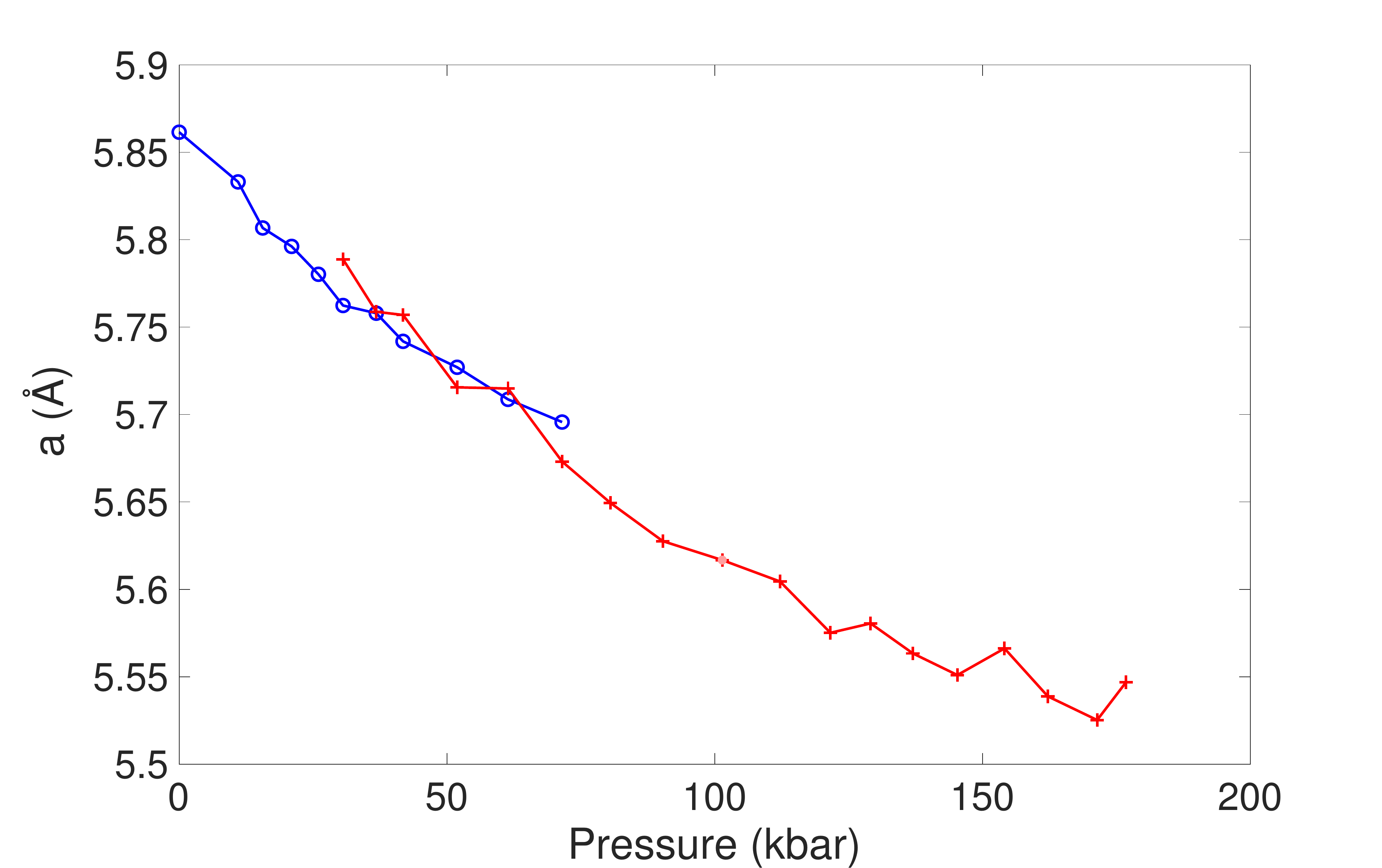}
\par\end{centering}
\begin{centering}
\includegraphics[width=0.5\columnwidth]{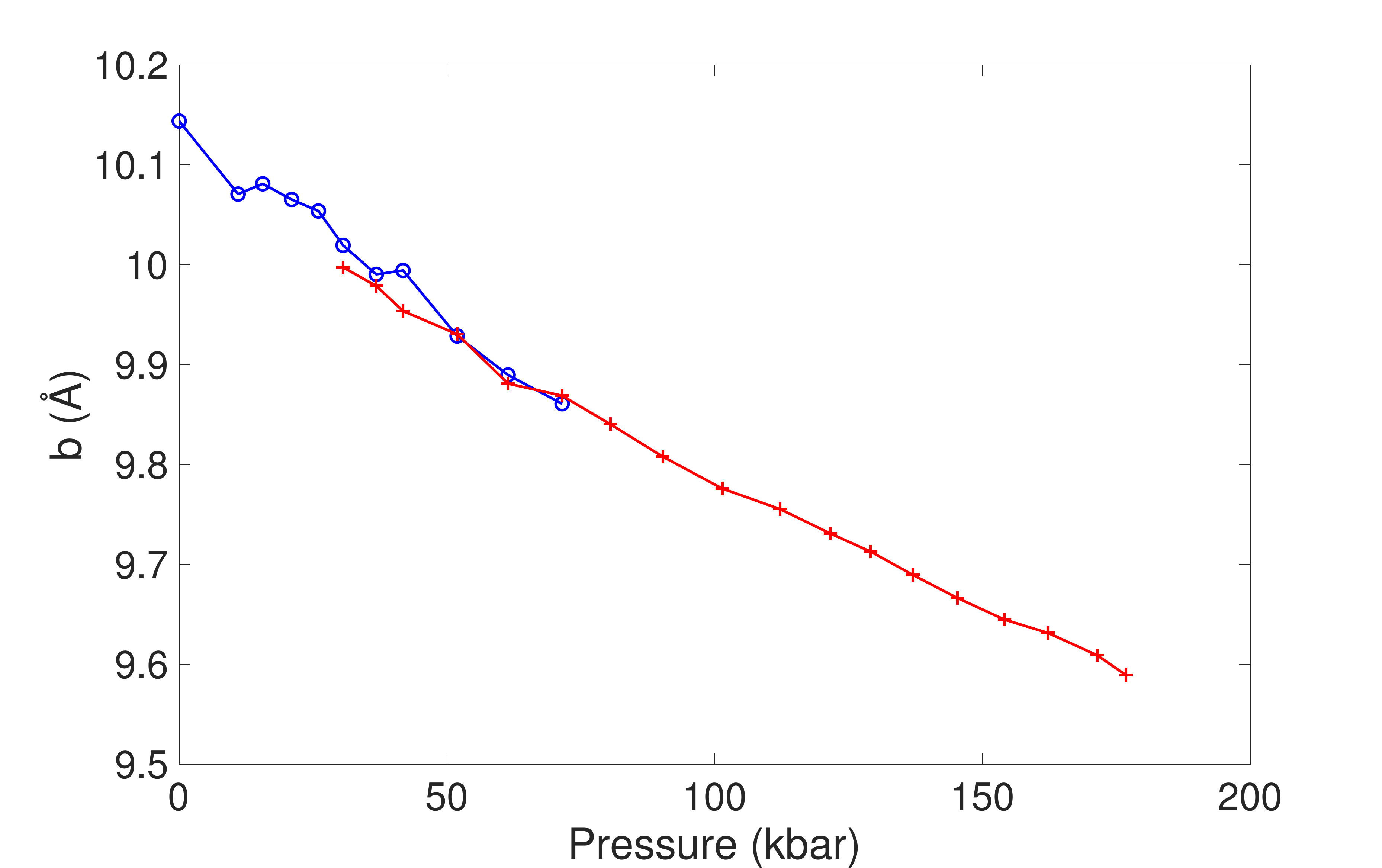}
\par\end{centering}
\centering{}\includegraphics[width=0.5\columnwidth]{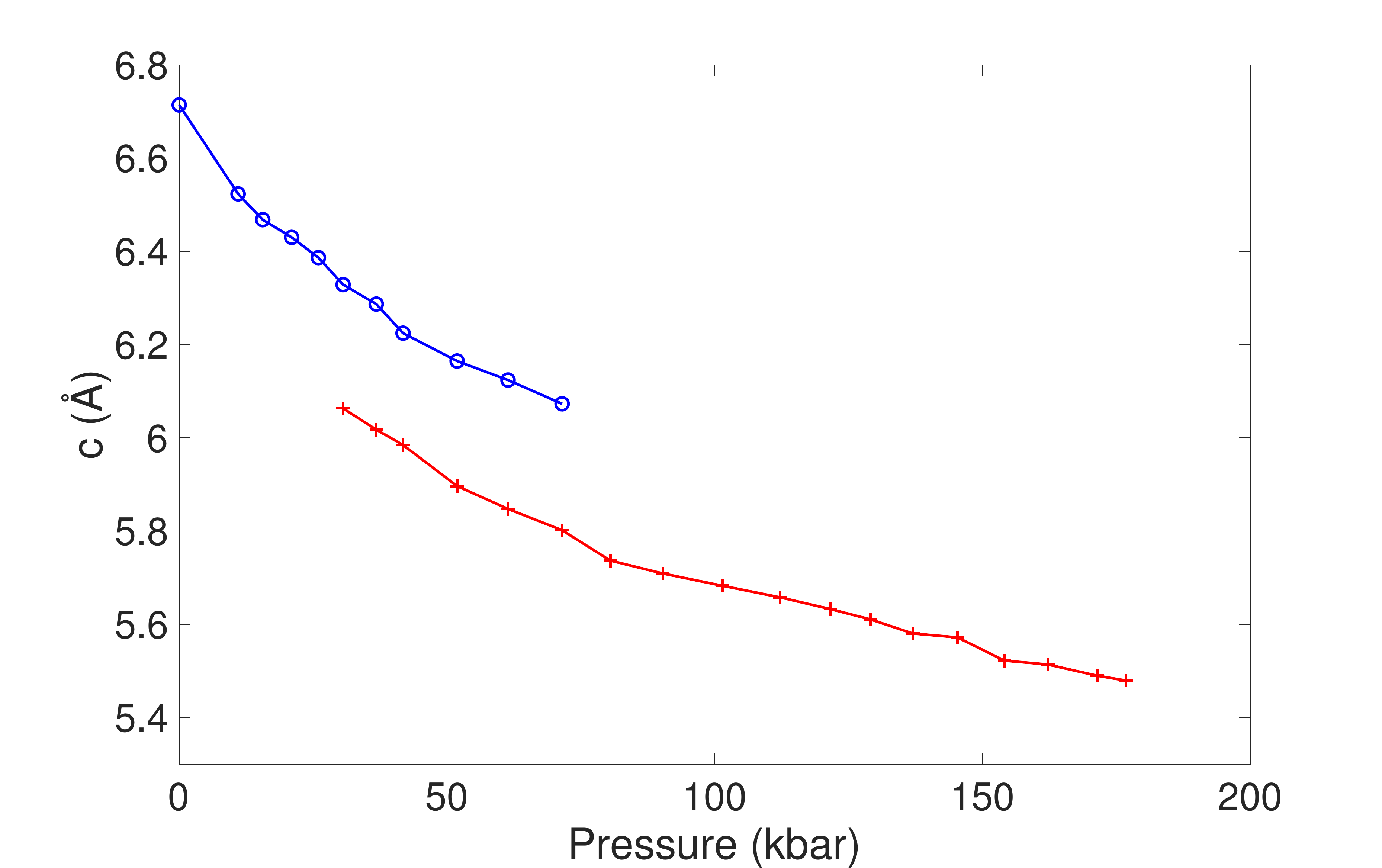}\caption{\label{fig:latticeParams}Lattice parameters extracted from powder
x-ray refinements for \vpps{} at room temperature as a function of
pressure. Blue circles denote the LP phase and red crosses the HP-I
phase. Note that while there is a jump in the $c$ axis value between
the two phases, this is due to a change in $\beta$ and does not correspond
to a change in inter-planar spacing or cell volume.}
\end{figure}

\begin{figure}[H]
\begin{raggedright}
\qquad{}a)
\par\end{raggedright}
\begin{centering}
\includegraphics[width=0.8\columnwidth]{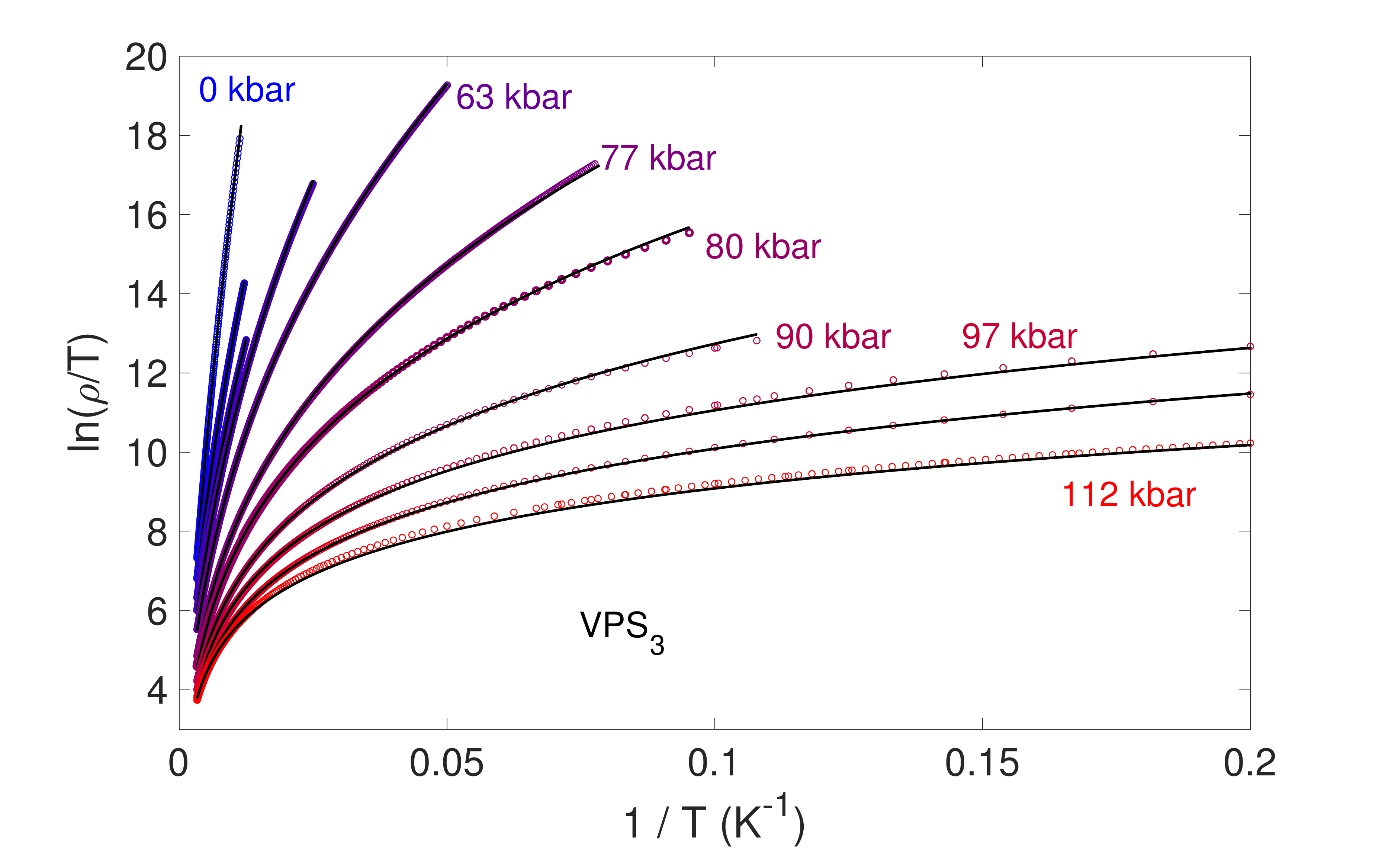}
\par\end{centering}
\begin{raggedright}
\qquad{}b)
\par\end{raggedright}
\centering{}\includegraphics[width=0.8\columnwidth]{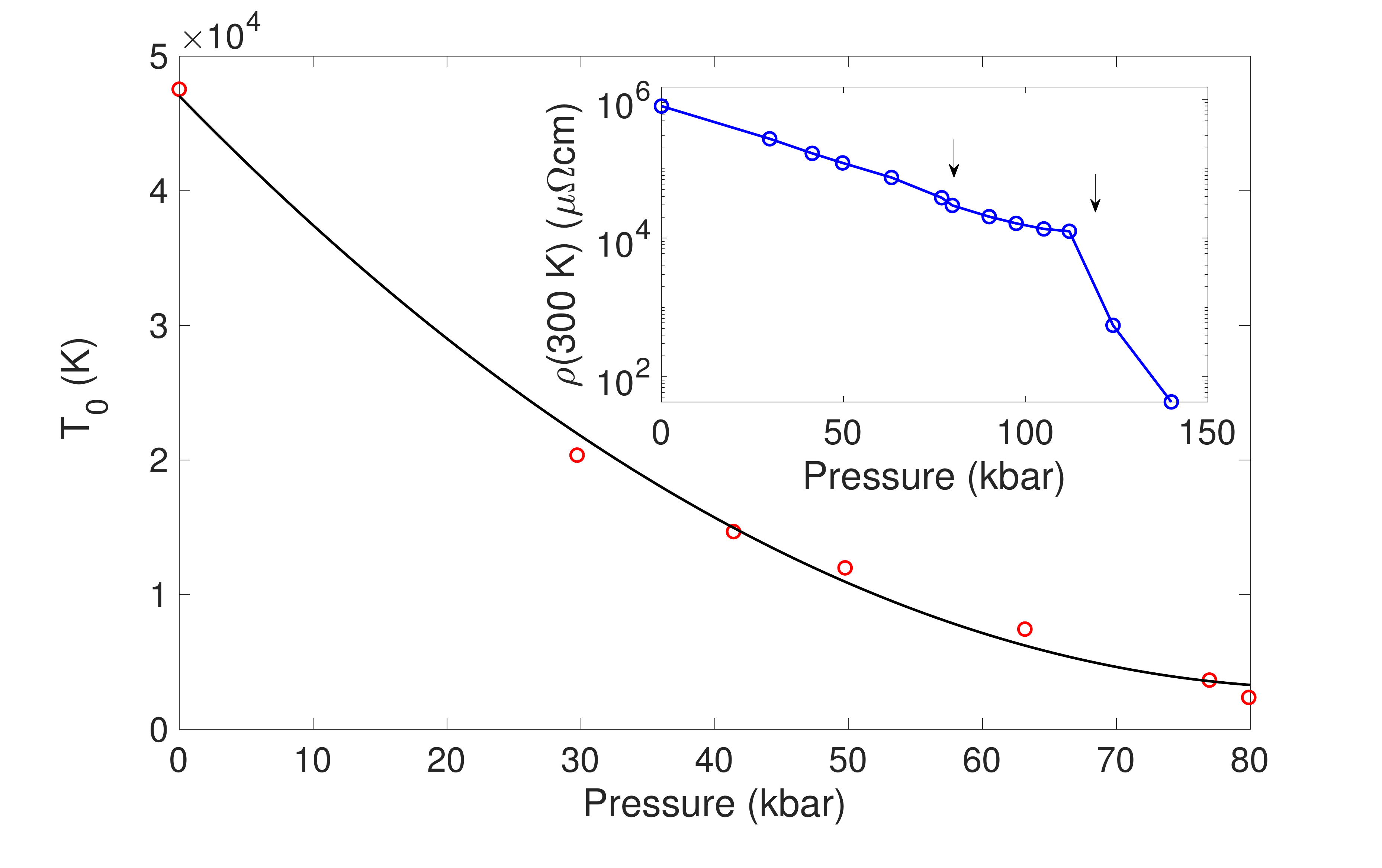}\caption{\label{fig:PowerLawFits}a) - $\mathrm{ln}(\rho/T)$ vs $1/T$ plotted
to extract the $\alpha$ power law dependence in the generalized variable
range hopping expression discussed in the main text $\rho=\rho_{0}Te^{(T_{0}/T)^{\alpha}}$.
Fits of type $1/T^{\alpha}$ are shown as black lines, allowing the
exponent $\alpha$ to be extracted. b) - extracted values of the $T_{0}$
characteristic temperature. Inset shows values of the room-temperature
resistivity as pressure is increased, on a logarithmic axis. Arrows
denote a kink or change of slope at 80~kbar and the insulator-metal
transition between 112-124~kbar.}
\end{figure}

Unlike in other \mpx{} materials such as \feps{} \citep{Haines2018b},
the resistivity cannot be well described by a simple Arrhenius-type
insulating temperature dependence - Fig. \ref{fig:PowerLawFits}.a
illustrates this by plotting $\mathrm{ln}(\rho/T)$ against $1/T$
- the resulting plots are far from straight lines and can be fitted
with a simple power law dependence. The simpler form $\mathrm{ln}(\rho)$
against $1/T$ for an equivalent plot gives an extremely similar result
and is the standard test for Arrhenius $e^{E_{a}/k_{b}T}$ resistivity.
Fig. \ref{fig:PowerLawFits}.b shows the extracted characteristic
temperature $T_{0}$ and the room temperature resistivity. If the
data follows variable-range-hopping expression $\rho=\rho_{0}Te^{(T_{0}/T)^{\alpha}}$,
plotting $\mathrm{ln}(\rho/T)$ against $1/T$ as in Fig. \ref{fig:PowerLawFits}.a
then yields an expression $\mathrm{ln}(\rho/T)=\mathrm{ln(}\rho_{0})+T_{0}^{\alpha}(1/T)^{\alpha}$
and so the exponent $\alpha$ can easily be found from a non-linear
least-squares fit. Such fits are shown overlayed onto the data curves
- good fits are seen up to pressures around 80~kbar, but then the
quality of fits starts to decrease as the data move away from the
VRH form, particularly at lower temperatures, as the insulator-metal
transition is approached.

As the transition is approached, the $T_{o}$ characteristic temperature
is continuously suppressed - again, no clear abrupt changes - until
the VRH expression breaks down in the close proximity of metallization
above 80~kbar. As electron overlap is increased by pressure, the
hopping energies are gradually decreasing. The decrease in the room-temperature
resistivity - a metric which is of course independent of any fitting
methodology - is once again smooth and continuous, except for the
insulator-metal transition at 112-124~kbar and a slight kink or change
of slope around 80~kbar. 80~kbar corresponds to the maximum pressure
of the LP - HP-I structural crossover, so a change in $\rho(p)$ can
be expected here as phase ratios are no longer changing and affecting
the resistivity.
\end{document}